\documentclass[fleqn,usenatbib]{mnras}

\usepackage{newtxtext,newtxmath}
\usepackage[T1]{fontenc}
\usepackage{ae,aecompl}

\usepackage{natbib}

\usepackage{etoolbox}
\makeatletter
\patchcmd\@combinedblfloats{\box\@outputbox}{\unvbox\@outputbox}{}{\errmessage{\noexpand patch failed}}
\makeatother

\usepackage{graphicx}	
\usepackage{amsmath}	
\usepackage{amssymb}	
\usepackage{xcolor} 	

\usepackage{hyperref}

\title[C-EAGLE mass comparison]{The Cluster-EAGLE project: a comparison of dynamical mass estimators using simulated clusters}

\author[T. J. Armitage et al.]{Thomas J. Armitage,$^{1}$\thanks{E-mail: thomas.armitage-3@postgrad.manchester.ac.uk}
	Scott T. Kay,$^{1}$
	David J. Barnes,$^{2}$
	Yannick M. Bah\'e$^{3}$ and
	\newauthor
	Claudio Dalla Vecchia$^{4,5}$
	\\
	$^{1}$Jodrell Bank Centre for Astrophysics, School of Physics and Astronomy, The University of Manchester, Manchester M13 9PL, UK\\
	$^{2}$Department of Physics, Kavli Institute for Astrophysics and Space Research, Massachusetts Institute of Technology, Cambridge, MA 02139, USA\\
	$^{3}$Leiden Observatory, Leiden University, PO Box 9513, NL-2300 RA Leiden, the Netherlands\\
	$^{4}$Instituto de Astrof\'\i{}sica de Canarias, E-38205 La Laguna, Tenerife, Spain\\
	$^{5}$Universidad de La Laguna, Dpto. Astrof\'\i{}sica, E-38206 La Laguna, Tenerife, Spain\\
}

\date{Accepted XXX. Received YYY; in original form ZZZ}

\pubyear{2018}

\begin{document}
	\label{firstpage}
	\pagerange{\pageref{firstpage}--\pageref{lastpage}}
	\maketitle
	
	\begin{abstract}
		Forthcoming large-scale spectroscopic surveys will soon provide data on thousands of galaxy clusters. It is important that the systematics of the various mass estimation techniques are well understood and calibrated. We compare three different dynamical mass estimators using the C-EAGLE galaxy clusters, a set of high resolution simulations with resolved galaxies \textcolor{black}{a median total mass, $M_{200c} = 10^{14.7} \, \mathrm{M_\odot}$}. We quantify the bias and scatter of the Jeans, virial, and caustic mass estimators using \textcolor{black}{all} galaxies with a stellar mass $M_*> 10^9 \, \mathrm{M_\odot}$, both in the ideal 3D case and in the more realistic projected case. On average we find our mass estimates are unbiased, though relative to the true mass within $r_{200c}$ the scatter is large with a range of $0.09$ - $0.15$ dex. We see a slight increase in the scatter when projecting the clusters. Selecting galaxies using the same criteria, we find no significant difference in the mass bias or scatter when comparing results from hydrodynamical and dark matter only simulations. However, selecting galaxies by stellar mass reduces the bias compared to selecting by total mass. Comparing X-ray derived hydrostatic and dynamical masses, the former are ${\sim} 30$ per cent lower. We find a slight dependence between substructure, measured using two different metrics, and mass bias. In conclusion, we find that dynamical mass estimators, when averaged together, are unbiased with a scatter of $0.11 \pm 0.02$ dex when including interloper galaxies and with no prior knowledge of $r_{200c}$.
	\end{abstract}
	
	
	
	
	\section{Introduction}
	Galaxy clusters form from the largest primordial density perturbations to have collapsed by the current epoch. As they trace the high mass tail of the halo mass function they are powerful cosmological probes (see \citealt{Allen2011,Kravtsov2012,Weinberg2013,Mantz2014}). However, to become precision probes of cosmology we require accurate and robust cluster mass estimates. Traditionally there are two major constraints on the usefulness of galaxy clusters as cosmological probes: understanding the astrophysical processes inside the clusters and acquiring enough high quality data to study them. Impending large scale surveys such as eBOSS, DESI, eROSITA, \textit{Euclid} and SPT-3G will increase the number of known clusters significantly, with \textit{Euclid} alone expected to find ${\sim} 10^6$ clusters with $M_{200c} > 10^{14} \, \rm M_\odot$\footnote{We define $M_{\rm 200c}$ as the mass enclosed within a sphere of radius $r_{\rm 200c}$ whose mean density is $200$ times the critical density of the Universe.} \citep{Laureijs2011}. Therefore, it is critical to understand the systematic limitations of cluster mass estimates to realise the full potential of these upcoming surveys.
	
	There are three primary methodologies used to estimate a galaxy cluster's mass: gravitational lensing, X-ray observations, and a dynamical analysis of cluster galaxies. Each requires their own set of assumptions and are limited by different systematics. For instance, the diffuse X-ray emission from the intracluster medium (ICM) can be affected by non-thermal pressure sources such as gas accretion, active galactic nuclei (AGN) feedback, substructures, turbulence and cosmic rays. This can lead to X-ray mass biases of $10-40$ per cent (\citealt{Lau2009,Rasia2012,Nelson2014,Henson2017}). Weak lensing estimates are independent of the dynamical state of the cluster, but are subject to projection effects, partly due to clusters being triaxial, with biases in dark matter only (DMO) simulations ${\sim} 5$ per cent \citep{Okabe2010,Oguri2011,Becker2011,Bahe2012,Mahdavi2013,Hoekstra2015,Kettula2015,Henson2017}. Dynamical mass estimates of clusters use the motions of galaxies inside the cluster as dynamical tracers of the gravitational potential. Studies of the velocity dispersion - mass relation have found that galaxies are not ideal tracers, with the velocity dispersion dependent on the galaxy sample selection. The bias has been reported to be $\pm 10$ per cent with respect to the underlying dark matter (DM) velocity dispersion (e.g. \citealt{Munari2013,Armitage2018,Elahi2018}). However, in a previous paper using the same simulations as in the current study, \cite{Armitage2018} found that stellar mass-limited galaxy samples yielded unbiased estimates of the DM velocity dispersion.
	
	\textcolor{black}{In addition one can use mass proxies, such as the Sunyaev-Zel'dovich (SZ) flux (e.g. \citealt{Planck2014XX,Saliwanchik2015}), optical richness (e.g. \citealt{Yee2003,Simet2017}), and the velocity dispersion, $\sigma$, of member galaxies (e.g. \citealt{Zhang2011,Bocquet2015,Sereno2015}) once they have been calibrated. While in principle the velocity dispersion is compelling mass proxy, which is insensitive to the assumed cosmology, with tight scatter \citep{Evrard2008}, there are several complicating factors. \cite{Biviano2006} found the Virial mass estimator to be biased high by 15 per cent, while the velocity dispersion - mass relation under-predicted the mass by 15 per cent for sample sizes greater than 60. \cite{Munari2013} and \cite{Armitage2018} only consider the ideal case where the full 3D velocity components are known. If instead one is limited to line-of-sight velocity measurements \cite{Saro2013} found that the intrinsic scatter of the measured velocity dispersion increases by approximately a factor of three from ${\sim} 13$ per cent up to ${\sim} 30 - 40$ per cent. \cite{White2010} found that the variance in the line-of-sight velocity dispersion was correlated with the orientation of the large scale structure surrounding the cluster.}
	
	In this paper we will consider the three main dynamical mass estimators: the virial, Jeans and caustic methods, all of which rely on spectroscopic observations. Multiple comparisons of observational data have already been made using clusters with both X-ray and spectroscopic observations for a few objects (e.g. \citealt{Diaferio2005,Rines2016,Maughan2016,Foex2017MC}). For example, \cite{Maughan2016} found that X-ray masses were ${\sim} 20$ per cent larger than the caustic masses. However, their value for the caustic filling factor, which is typically constrained using simulations, $\mathcal{F}_\beta=0.5$, may be the main cause of this difference as $\mathcal{F}_\beta$ has been found by different authors to be between $0.5-0.7$ depending on the simulation and the desired radius to measure the mass \citep{Diaferio1999,Serra2011,Gifford2013}. \cite{Foex2017MC} studied 10 galaxy clusters with hundreds of spectroscopically-measured galaxies per object. They compared dynamical mass estimates for Jeans, caustic and virial methods, finding the masses to be ${\sim} 20$, ${\sim} 30$ and ${\sim} 50$ per cent higher than the X-ray masses respectively. \cite{Foex2017MC} found that by excluding galaxies thought to be part of substructure, any statistically significant difference between the three dynamical masses and the X-ray mass could be eliminated.
	
	Simulations of clusters play an important role as they allow us to determine the absolute bias between different mass estimators and the true value (e.g. \citealt{Rasia2006,Nagai2007,Lau2009,Serra2011,Rasia2012,Gifford2013,Gifford2013b,Nelson2014,Old2014,Caldwell2016,Gifford2017}), \textcolor{black}{as well as the relative difference between techniques}. Hydrodynamical simulations, as opposed to semi-analytic models, are the only way to self-consistently model the baryonic effects during cluster formation. However, a limitation has traditionally been numerical resolution, with a typical gas particle mass of ${\sim} 10^9 \, \mathrm{M_\odot}$ and spatial resolution ${\sim} 5 \, \mathrm{kpc}$ for cosmological simulations (e.g. \citealt{Planelles2013,LeBrun2014, Pike2014, Bocquet2016, Barnes2017,McCarthy2017}). The lack of resolution can result in the failure to capture dynamical processes, such as preferential stripping of DM relative to stars in infalling galaxies \citep{Smith2016}. In the last few years there has been a leap forward in the numerical resolution of cosmological hydrodynamical simulations. One such set of simulations is the EAGLE suite \citep{Schaye2015,Crain2015}, which has a gas particle mass of $1.8\times 10^6 \, \mathrm{M_\odot}$ (for the reference box), sufficient to resolve galactic structure. Due to the computational expense, the largest simulated volume is only $100 \, \mathrm{Mpc}$ on a side, too small for many clusters to form. The Cluster-EAGLE (C-EAGLE) simulations \citep{Barnes2017b,Bahe2017}, consist of 30 high resolution, hydrodynamical galaxy clusters simulated using the EAGLE subgrid physics applied to a set of zoom simulations (e.g. \citealt{Tormen1997}). The C-EAGLE clusters are arguably the first to resolve realistic galactic structure in the cluster environment, and capture dynamical process that would otherwise be missed in lower resolution simulations \citep{Bahe2017,Armitage2018}. We use the C-EAGLE clusters to test the virial, Jeans and caustic mass estimators, both in the ideal case with known 3D galaxy positions and velocities, as well as the more realistic case with line of sight (LoS) quantities. We use the 3D scenario to quantify how robust the mass estimators are in the best possible case and as a point of reference to understand the importance of projection effects and interlopers in the 2D analysis.
	
	The paper is organised as follows. In Section \ref{sims} we give a brief overview of the C-EAGLE simulations, while Section \ref{tech} describes the mass estimation techniques used in this paper and the assumptions made in their implementation. We then present our mass estimation results in Section \ref{res}. Section \ref{sub} describes the processes used to identify and quantify the presence of dynamical substructures, before presenting how substructure is correlated with mass bias. Finally, we conclude our findings in Section \ref{conc}.

	\section{C-EAGLE simulations} \label{sims}
	Here, we briefly summarise the C-EAGLE sample used in this paper. For a more detailed description of the C-EAGLE dataset and subgrid model, see \cite{Barnes2017b} and \cite{Bahe2017}.
	
	The C-EAGLE clusters comprise of 30 zooms (labelled CE-00 - CE-29, with the higher numbers approximately corresponding to more massive clusters) spanning 10 logarithmically-spaced mass bins between $14.0 \le \log (M_{\rm 200c}/\mathrm{M_\odot}) \le 15.4$ at redshift zero\footnote{\textcolor{black}{The true value of $M_{200c}$ is found by summing the mass of all particles within a sphere of radius $r_{200c}$, centred on the particle with the most negative gravitational potential.}}, selected from a large $(3.2 \, \mathrm{Gpc})^3$ parent simulation\footnote{We use $\log$ to refer to $\log_{10}$ and $\ln$ refers to the natural logarithm.}. The initial gas particle mass for the C-EAGLE sample is $m_\mathrm{gas}=1.8 \times 10^6 \, \mathrm{M_\odot}$ and the DM particle mass is $m_\mathrm{DM} = 9.7 \times 10^6 \, \mathrm{M_\odot}$. The gravitational softening length was set to $2.66$ co-moving $\mathrm{kpc}$ until $z=2.8$ and $0.70$ physical $\mathrm{kpc}$ at lower redshift. The underlying cosmology assumed was $\Lambda\mathrm{CDM}$, based on the \textit{Planck} 2013 results combined with baryonic acoustic oscillations, WMAP polarization and high multipole moments experiments \citep{Planck2014I}. The cosmological parameters were set to $\Omega_{\rm{b}}=0.04825$, $\Omega_{\rm{m}}=0.307$, $\Omega_{\Lambda}=0.693$, $h\equiv H_0/(100\,\rm{km}\,\rm{s}^{-1}\,\rm{Mpc}^{-1})=0.6777$, $\sigma_{8}=0.8288$, $n_{\rm{s}}=0.9611$ and $Y=0.248$. The high resolution region of each cluster extends to at least $5 r_{\rm 200c}$ before any contaminating low resolution particles are encountered. For the purposes of this paper, we ignore all particles beyond $5 r_{\rm 200c}$.
	
	The C-EAGLE clusters were run using the same code as the EAGLE simulations \citep{Schaye2015,Crain2015}. This code is based upon a modified version of the \textit{N}-Body Tree-PM SPH code \textsc{P-Gadget-3}, last described in \cite{Springel2005}. The implemented hydrodynamics is collectively known as \textsc{anarchy} (for details see Appendix A of \citealt{Schaye2015} and \citealt{Schaller2015}). \textsc{anarchy} is based on the pressure-entropy formalism derived by \cite{Hopkins2013} with an artificial viscosity switch \citep{CullenDehnen2010} and includes artificial conductivity similar to that suggested by \cite{Price2008}. The $C^2$ smoothing kernel of \cite{Wendland1995} and the time-step limiter of \cite{DurierDallaVecchia2012} are also used.
	
	The EAGLE code is based on that of the OWLS project \citep{Schaye2010}, also used in the GIMIC \citep{Crain2009} and COSMO-OWLS \citep{LeBrun2014} simulations. This includes radiative cooling, star formation, stellar feedback and the seeding, growth and feedback of black holes \citep{SchayeDallaVecchia2008,WiersmaSchayeSmith2009,DallaVecchia2012,RosasGuevara2015}. For the EAGLE code the effects of star formation and feedback were calibrated to reproduce a limited set of observational data. \cite{Schaye2015} presented three calibrated subgrid models that matched observations of the galaxy stellar mass function and galaxy mass-size relation, REF, AGNdT9 and Recal. The Recal model is not relevant as that is intended for a mass resolution $8\times$ greater than the standard EAGLE simulations. The main difference between the REF and AGNdT9 models is the heating temperature, $\Delta T$, which is $10^{8.5} \, \rm K$ and $10^{9} \, \rm K$ for the REF and AGNdT9 models respectively. The second difference is an increase in the effective viscosity around the subgrid accretion disk of black holes by a factor of $10^2$. The AGNdT9 model presented in \cite{Schaye2015} is a better match to the observed X-ray luminosities and gas mass fractions of low mass groups ($M_{500c}<10^{13.5} \, M_\odot$) present in the simulation volume. It was for this reason that the AGNdT9 subgrid model was chosen for C-EAGLE.
	
\begin{figure}
	\centering
	\includegraphics[width=0.99\linewidth]{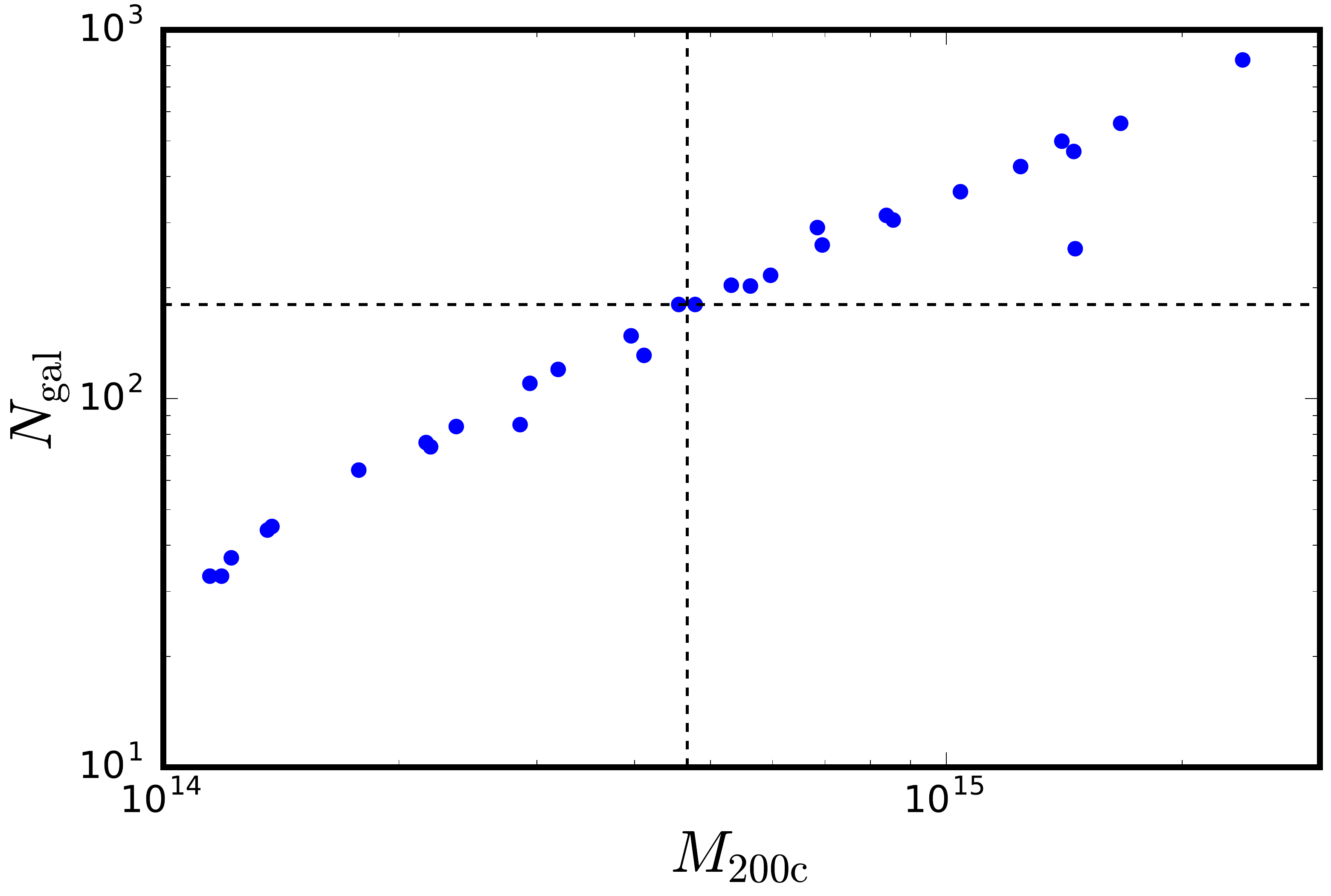}
	\caption{\color{black} The number of galaxies, $N_\mathrm{gal}$, with stellar mass $M_*>10^9 \, \rm M_\odot$ inside $r_{200c}$ for each cluster as a function of $M_{200c}$. The vertical and horizontal dashed lines mark the median $M_{200c}$ and $N_\mathrm{gal}$ respectively.}
	\label{fig:MassRich}
	\end{figure}

	In this paper we define any self-bound object, as determined by the SUBFIND algorithm \citep{Springel2001,Dolag2009}, with a stellar mass greater than $10^9 \, \mathrm{M_\odot}$ as a galaxy and use it in the subsequent analysis. \textcolor{black}{Fig. \ref{fig:MassRich} shows the number of galaxies in each cluster against the total mass. The median number of galaxies in a cluster is 180, marked by the horizontal dashed line. Our sample is intended to represent a high quality dataset that would be used as a reference for other surveys.}
	
	We consider two scenarios; the `ideal' case where we use the true 3D values for the galaxy positions and velocities, as well as the more realistic case using line of sight (LoS) quantities, ($x,y,\varv_z$). For the LoS case we project the cluster down a cylinder of length $10 r_\mathrm{200c}$, centred on the cluster centre of potential, and we do not attempt to remove interloper galaxies, as we found that using the shifting-gapper technique (\citealt{Fadda1996,Gifford2013}) introduced a bias in the velocity dispersion. \textcolor{black}{(We note that the relatively high mass of the C-EAGLE clusters also limits the impact of interloper galaxies.)} The radius of the cylinder is at least $1.5 r_{200c}$. As all galaxies lie within $5r_{200c}$ of the cluster, this already represents a well selected sample and is not necessarily representative of the performance of the shifting-gapper technique using lightcones to make mock observations. We tested how sensitive our results were to the $5r_{200c}$ cut, using the 13 \textit{Hydrangea} clusters in our sample, where the high resolution volume extends to $10r_{200c}$ \citep{Bahe2017}. We found no significant increase in the bias or scatter when taking galaxies from within $5$ or $10r_{200c}$. The `ideal' case is to demonstrate the upper limit of what could be achieved using galaxies as tracers, whereas the latter case shows a more realistic scenario. All of the analysis is performed at redshift zero.
	
	\section{Mass estimation techniques} \label{tech}
	We will now discuss the mass estimation techniques used in this paper. The following subsections detail the three methods used in this paper in turn, including the justifications for the required assumptions.
	
	\subsection{Caustic method}
	\begin{figure}
		\centering
		\includegraphics[width=0.99\linewidth]{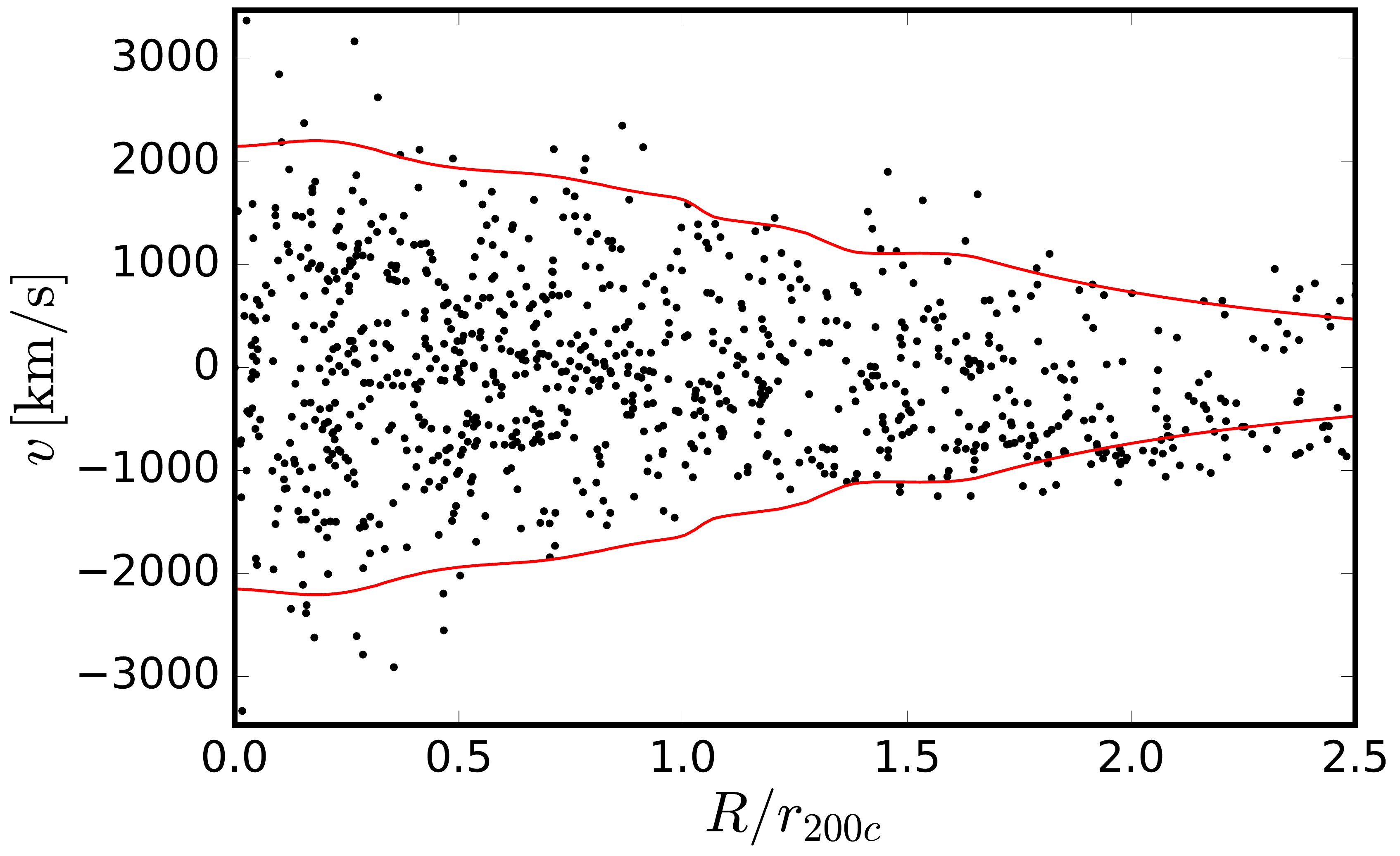}
		\caption{An example caustic profile for CE-26. The black points are galaxies with a stellar mass greater than $10^9$ $\mathrm{M_\odot}$ within a cylinder of length $10r_{200c}$. The solid red line is the caustic profile, obtained from the galaxies' line of sight velocities, showing the estimated escape velocity of the cluster. Both the galaxy velocities and radial separation are projected quantities. The true $M_{200c}$ of this cluster is $1.45\times10^{15} \, \mathrm{M_\odot}$, whereas the estimated mass inside $R(r_{200c})=2.39 \,\mathrm{Mpc}$ is $1.56\times10^{15} \mathrm{M_\odot}$.}
		\label{fig:ExampleCaustic}
	\end{figure}
	The caustic method does not rely upon an assumption that the cluster is in dynamical equilibrium. Instead, the method, first proposed by \citeauthor{Diaferio1997} (\citeyear{Diaferio1997}; see also \citealt{Diaferio1999}), attempts to measure the escape velocity of the galaxy cluster as a function of radius. It does this by noting that any object with a speed greater than the escape velocity, $\varv_\mathrm{esc}$, of the cluster will not reside within the cluster for long. As the escape velocity is directly related to the cluster potential, $\varv^2_{\mathrm{esc}}(r)=-2\phi(r)$, we can obtain the integrated mass profile $M(r)$ by measuring the escape velocity as a function of radius. Accounting for the fact that only the galaxy line of sight velocities can be measured, \cite{Diaferio1997} found that the mass profile of a cluster can be written as
	\begin{equation}
	GM(r)=\int_{0}^{r} \mathcal{A}^2(s) \mathcal{F}_{\beta}(s) ~ \mathrm{d}s  \:,
	\end{equation}
	where $\mathcal{A}$ is the caustic amplitude and is a measure of the escape velocity as a function of radius. The filling factor $\mathcal{F}_\beta$ accounts for the projection along the line of sight
	\begin{equation}
	\mathcal{F}_{\beta} = -2 \pi G \frac{\rho(r)r^2}{\phi(r)} \left( \frac{3-2\beta(r)}{1-\beta(r)} \right) \:,
	\end{equation}
	where $\rho$ is the cluster mass density profile and $\beta$ is the velocity anisotropy, defined as
	\begin{equation}
	\beta=1-\frac{\sigma^2_\phi + \sigma^2_\theta}{2 \sigma^2_r} \:,
	\end{equation}
	where $\sigma$ is the velocity dispersion measured in spherical polar coordinates ($r,\theta,\phi$). 
	
	The big step taken by \cite{Diaferio1997} was to state that $\mathcal{F}_\beta$ is approximately constant with radius, with choices in the literature varying from 0.5 to 0.7 (\citealt{Diaferio1997,Diaferio1999,Serra2011,Gifford2013,Gifford2013b}). In this work we take $\mathcal{F}_\beta=0.75$, chosen to minimise the bias in the 3D case, which we then applied to our contaminated 2D sample. Our value of $\mathcal{F}_\beta$ is slightly higher than in other literature. This is primarily due to us only considering the value of $M_{200c}$ in the calibration, as opposed to the whole mass profile. Because $\mathcal{F}_\beta$ is not in practice constant with radius, the desired radially averaged value will differ depending on the radial range of interest, which in our case is a single point.
	
	Our implementation of the caustic method is based on that of \cite{Gifford2013}\footnote{Their code is publicly available at \url{https://github.com/giffordw/CausticMass}}. Summarising the method, the caustic amplitude $\mathcal{A}(r)$ is found by identifying a density threshold in projected phase space $(\varv_\mathrm{LoS},R)$, after smoothing with an appropriate kernel, in our case a multidimensional Gaussian in $(\varv_\mathrm{LoS},R)$. \cite{Serra2011} enforce $\mathrm{d \ln}\varv_{\rm esc} / \mathrm{d \ln} r \leqslant 2$, which removes drastic, and likely unphysical, changes in the escape velocity while not being overly restrictive. If the gradient is exceeded, the escape velocity at that point is set so that $\mathrm{d \ln}\varv_{\rm esc} / \mathrm{d \ln} r = 2$. For more details see \cite{Gifford2013} and \cite{Gifford2013b}. Note that we use the traditional non-parametric model, with a fixed value of $\mathcal{F}_\beta$, throughout this paper. An example caustic profile is shown in Fig. \ref{fig:ExampleCaustic}.
	
	\subsection{Virial theorem} \label{sec:VirialTheory}
	The implementation of the virial theorem is based on that of \cite{Limber1960} and \cite{Heisler1985}. In the case where we have all three velocity components, the virial mass, $M_v$, is given by
	\begin{equation}
	M_v=\frac{3}{G}\sigma_{1D}^2 R_\mathrm{H} \:,
	\label{eqn:Virial_3D}
	\end{equation}
	where $\sigma_{1D}^2=(\sigma_x^2+\sigma_y^2+\sigma_z^2)/3$ and $R_\mathrm{H}$ is the mean harmonic radius
	\begin{equation}
	R_\mathrm{H}=\frac{N_\mathrm{gal}(N_\mathrm{gal}-1)}{\sum_{i<j} \frac{1}{R_{i,j}} }\:,
	\end{equation}
	where $R_{i,j}$ is the radial distance between any two galaxies and $N_\mathrm{gal}$ the number of galaxies in the sample. When projected quantities are used equation (\ref{eqn:Virial_3D}) must be modified to
	\begin{equation}
	M_v=\frac{3 \pi}{2 G}\sigma_\mathrm{proj}^2 R_\mathrm{H}\:,
	\label{eqn:Virial_los}
	\end{equation} 
	where $\sigma_\mathrm{proj}$ is the line of sight velocity dispersion. The $\pi/2$ difference arises from changing between projected separations and true 3D separations, as shown in \cite{Limber1960}. The velocity dispersion is calculated using the gapper method, which was found by \cite{Beers1990} to be robust down to as few as 5 members. For this method, the velocities, $\varv$, of the $N_\mathrm{gal}$ galaxies are first sorted in increasing size. The velocity dispersion is then calculated using
	\begin{equation}
	\sigma_{\mathrm{gap}}=\frac{\sqrt{\pi}}{N_\mathrm{gal}(N_\mathrm{gal}-1)} \sum_{i=1}^{N_\mathrm{gal}-1} i(N_\mathrm{gal}-i)(\varv_{i+1} - \varv_i) \:.
	\label{sGapper}
	\end{equation}

	The virial theorem is, in essence, a simplified form of Jeans analysis, which relies upon the cluster being spherical and in dynamical equilibrium. It also relies upon the galaxies being fair tracers of the underlying DM particles, which is known to not necessarily be the case (e.g. \citealt{Biviano2006,Munari2013,Armitage2018}). Although the virial theorem does not return a mass profile, its simplicity still makes it an attractive mass estimator (e.g. \citealt{Biviano2006,Foex2017MC}).
	
	The above equations implicitly assume that the cluster is completely isolated from the rest of the Universe. In reality, clusters exist in a dense environment, continually accreting matter from their surroundings. In order to account for this, an additional term must be included when calculating the virial mass. This surface pressure term (SPT) corrects for the dynamical pressure of material falling onto the cluster (\citealt{Binney1987,Carlberg1996}). The corrected mass, measured within a radius $b$, is then given by
	\begin{equation}
	M_{cv}(b)=M_v \left[ 1 - 4\pi b^3 \frac{\nu(b)}{\int_{0}^{b} 4 \pi r^2 \nu(r) \mathrm{d}r} \left( \frac{\sigma_r(b)}{\sigma(<b)} \right)^2  \right] \:,
	\label{eqn:SPTterm}
	\end{equation}
	where $\nu$ is the galaxy number density. The velocity dispersion term, $\sigma_r(b) / \sigma(<b)$, is at most 1/3 if one assumes an isotropic velocity dispersion that is decreasing with radius \citep{Foex2017MC}. Without the inclusion of the SPT, virial masses tend to be overestimated by ${\sim} 20$ per cent (\citealt{Carlberg1997,Girardi1998}). We find the SPT correction to be ${\sim} 25 \pm 10$ per cent for the C-EAGLE clusters.
	
	In order to calculate the virial radius, $r_{200c}$, we use the iterative method of \cite{Foex2017VI}, where $r_{200c}$ is given by
	\begin{equation}
	r_{200c}=\left( \frac{3 M_{cv}}{4 \pi 200 \rho_\mathrm{cr}(z)} \right)^{1/3} \:,
	\label{eqn:virRadius}
	\end{equation}
	where $\rho_\mathrm{cr}(z)$ is the critical density at redshift $z$. We first calculate the corrected virial mass using equation (\ref{eqn:SPTterm}) using a large aperture, typically $3 \, \mathrm{Mpc}$, and from that calculate the virial radius using equation (\ref{eqn:virRadius}). We then recompute the virial mass using only galaxies within the calculated $r_{200c}$, iterating until convergence, for both the 2D and 3D cases separately.

	\subsection{Jeans analysis: method}
	
	The Jeans equation, appropriate for dynamical equilibrium and spherical symmetry, is given by the first moment of the collisionless Boltzmann equation
	\begin{equation}
	M(r)=-\frac{\sigma_r^2 r^2}{G} \left( \frac{\mathrm{d}\ln \nu}{\mathrm{d}r} + 2 \frac{\mathrm{d}\ln \sigma_r}{\mathrm{d}r} +2\frac{\beta}{r} \right)  \:,
	\label{eq:JeansEq}
	\end{equation}
	where $M(r)$ is the enclosed mass within $r$, $\sigma_r$ the radial velocity dispersion, $\beta$ the velocity anisotropy and $\nu$ the density profile of a tracer population, i.e. galaxies. As equation (\ref{eq:JeansEq}) requires us to find the derivatives of both the density and velocity dispersion profiles, we can make our measurements more robust to the inherently noisy data by using parametric models. Here we describe the models used to fit each property as a function of radius and in Section \ref{JeansSyst} we discuss and justify our modelling assumptions.

	We fit both the density and velocity dispersion profiles by placing the galaxies into radial bins of width $0.1 r_{200c}$, with the innermost bin edge at $0.05 r_{200c}$. If there are fewer than 10 galaxies in a given bin then the bin width is extended outwards until it contains 10 galaxies; this is done out to $1.5r_{200c}$. The galaxy threshold is chosen to reduce statistical noise when fitting for the profile. We do not fit the profile beyond $1.5r_{200c}$ as the empirical models are likely to be a poor fit at extended radii and we are only interested in the region around $r_{200c}$ in this work. We take the uncertainty in each galaxy density bin to be the square root of the number of galaxies per bin. The uncertainty in the measured velocity dispersion was found by bootstrapping the galaxies in each bin 1,000 times, calculating the velocity dispersion of each sample and taking the standard deviation of that distribution to be the error.
	
	We fit a two parameter NFW density profile \citep{Navarro1997} to the tracer density profile
	\begin{equation}
	\nu_\mathrm{NFW}(r) = \frac{\nu_0}{\frac{r}{r_s} \left(1+\frac{r}{r_s} \right)^2} \:,
	\label{eq:tracerNFW}
	\end{equation}
	where $r_s$ is the scale radius and $\nu_0$ the normalisation. Note that the scale radius can be related to the virial radius $r_{200c}$ by $r_{200}=r_{s}c$, where $c$ is the concentration. In the case of projected data we fit a projected NFW profile instead. Using the best fit value of $r_s$, the gradient of $\ln \nu_\mathrm{NFW}$ can be calculated via
	\begin{equation}
	\frac{\mathrm{d} \ln \nu_\mathrm{NFW}}{\mathrm{d}r} = - \left( \frac{1}{r} + \frac{2}{r+r_s} \right) \:.
	\end{equation}
	To find a robust estimate of the velocity dispersion gradient, we use a simple power law of the form 
	\begin{equation}
	\sigma=\sigma_0(1+r)^p \:,
	\label{sigma_plaw}
	\end{equation}
	where $p$ is the power law index and $\sigma_0$ is the central dispersion \citep{Carlberg1997,Stark2017,Aguerri2017}. The logarithmic slope is then given by
	\begin{equation}
	\frac{\mathrm{d} \ln \sigma^2_\mathrm{PWR}}{\mathrm{d}r} =\frac{2p}{1+r} \:.
	\end{equation}
	In our projected sample, we must also map $\sigma_\mathrm{LoS}$ on to $\sigma_r$. We take the approximation, also applied in the caustic analysis, that $ \langle \varv_\phi^2 \rangle = \langle \varv_\theta^2 \rangle \approx \langle \varv_\mathrm{LoS}^2 \rangle$. Assuming little bulk rotation when averaging close to $r_{200c}$ then $\sigma_r^2=\langle \varv_r^2 \rangle$ and therefore
	\begin{equation}
	\sigma_r^2 = \frac{\sigma_\mathrm{LoS}^2}{1-\beta} \:.
	\label{eqn:betaCorr}
	\end{equation}
	
	The final component is the $\beta$ profile. This is very difficult to obtain observationally as it requires knowledge of the 3D galaxy velocities. Previous work has typically focused on obtaining $\beta$ profiles for a few clusters, whether that is through using mass profiles obtained with X-ray analysis \citep{Benatov2006,Hwang2008,Host2009}, or just using dynamical information \citep{Biviano2004,Lokas2006,Wojtak2010}. There are relatively few studies which attempt to obtain the $\beta$ profile for a large collection of clusters (\citealt{Host2009,Wojtak2010,Stark2017}). There has also been the development of the MAMPOSSt algorithm \citep{Mamon2013}, which uses a maximum likelihood estimator to fit models of the mass profile, $\beta$, $\sigma$ and $\nu$. In this work we use the true $\beta$ profile in the 3D case, or assume a constant profile of $\beta=0.36$ in the 2D sample, which we justify below in Section \ref{JeansSyst}. Once the velocity dispersion, number density and $\beta$ profiles have been obtained for a cluster we then fit the integrated mass profile, $M(r)$, assuming an NFW model
	\begin{equation} \label{eq:NFWMassProf}
	M(r) = 4 \pi \rho_0 r_s \left[ \ln \left(\frac{r_s + r}{r_s} \right) -\frac{r}{r_s+r} \right] \, ,
	\end{equation}
	where $r_s$ is the scale radius as before and $\rho_0$ is analogous to $\nu_0$. We then either take the mass at $r_{200c}$ if known a-priori or we find the ($M_{200c}, r_{200c}$) pair directly from the fitted profile. 
	
	\subsection{Jeans analysis: modelling} \label{JeansSyst}
	There are several key assumptions that we make in order to obtain the mass profile: the galaxy number density profile is well approximated by an NFW profile, likewise the velocity dispersion profile is approximated by equation (\ref{sigma_plaw}), the $\beta$ profile is approximately constant with radius, and the radial velocity dispersion can be recovered from the line of sight velocity dispersion via equation (\ref{eqn:betaCorr}). We now discuss each of these assumptions in turn.
	
\begin{figure}
		\centering
		\includegraphics[width=0.99\linewidth]{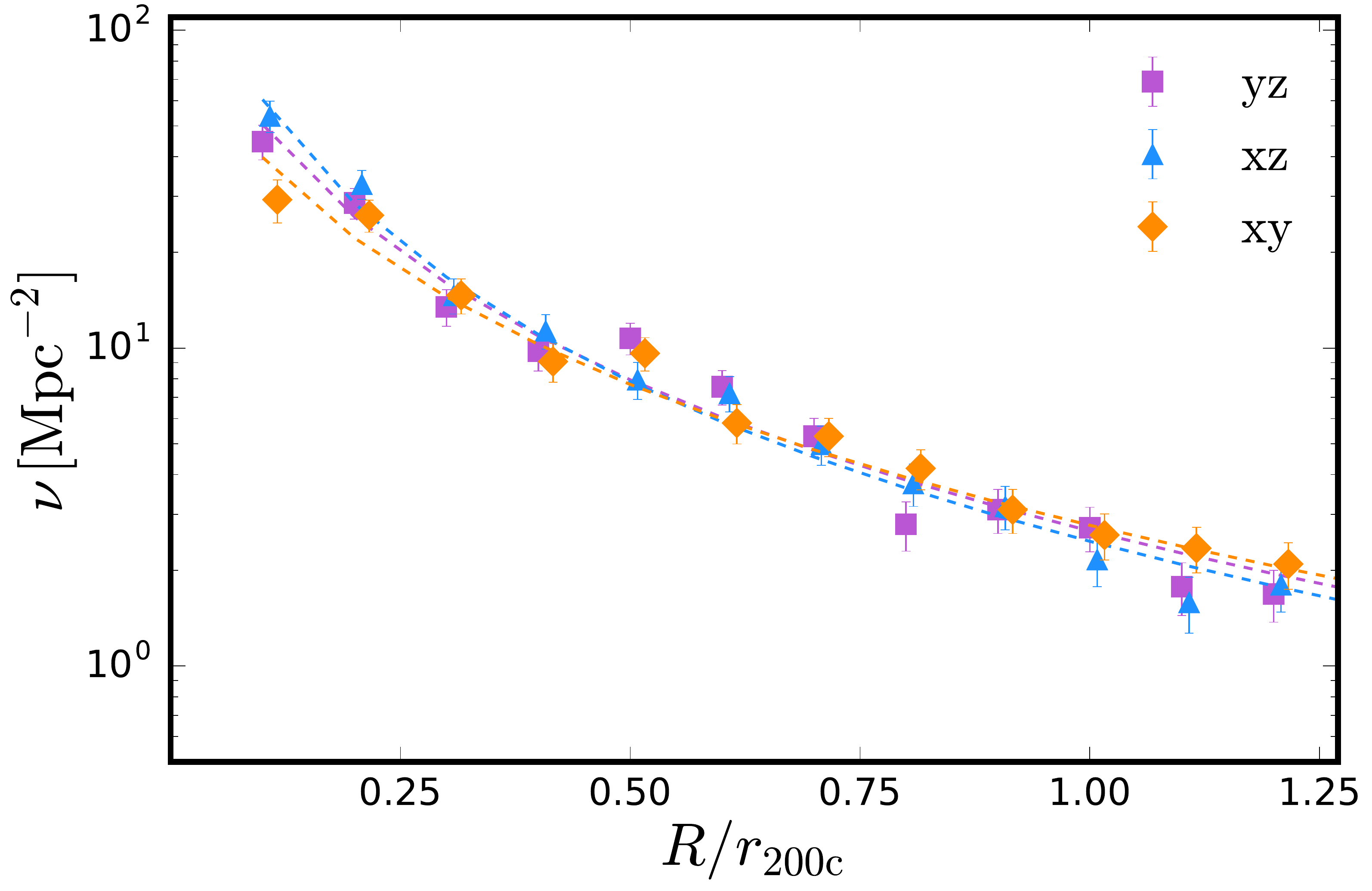}
		\caption{The projected surface density of CE-26 for three orthogonal projections, labelled according to the plane the galaxies were projected on to. The error bars denote the $1\sigma$ uncertainty on $\nu$ for each bin. Each profile is fit by a projected NFW model. Other profiles are shown in Fig. \ref{fig:SurfDens}.}
		\label{fig:NFW26}
	\end{figure}

	The galaxy number density profile is well approximated by an NFW model. We show CE-26 as an example in Fig. \ref{fig:NFW26} and the profiles for all clusters can be seen in Fig. \ref{fig:SurfDens}. The purple squares, blue triangles and orange diamonds show the profile recovered for each of the three orthogonal cluster projections used in this work. Both equation (\ref{eq:tracerNFW}) and equation (\ref{eq:NFWMassProf}) include the scale radius, $r_s$, as a fit parameter. As one is fitting the same underlying mass distribution the values of $r_s$ should be similar. We determine the optimum value for $r_s$ independently for the $\nu$ and $M(r)$ profiles allowing us to test if they differ substantially, which as shown in Fig. \ref{fig:RsComparison} they do. This implies that the galaxies are not fair tracers of the underlying density profile, and justifies the assumption that the two $r_s$ values should be fit independently. I.e. even though both the mass and number density profiles are both fit well by an NFW model the fit parameters differ.
	
\begin{figure}
		\centering
		\includegraphics[width=0.99\linewidth]{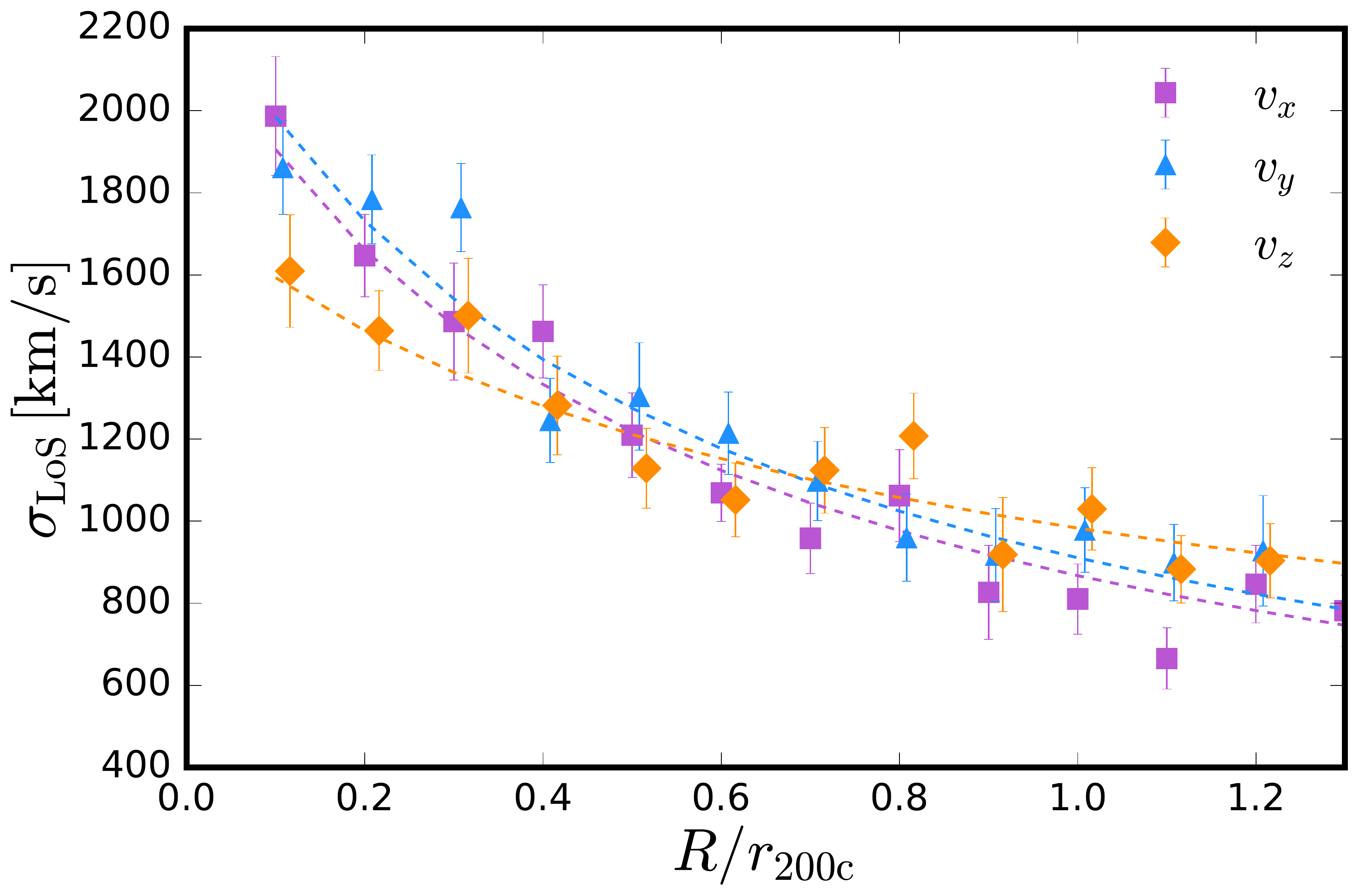}
		\includegraphics[width=0.99\linewidth]{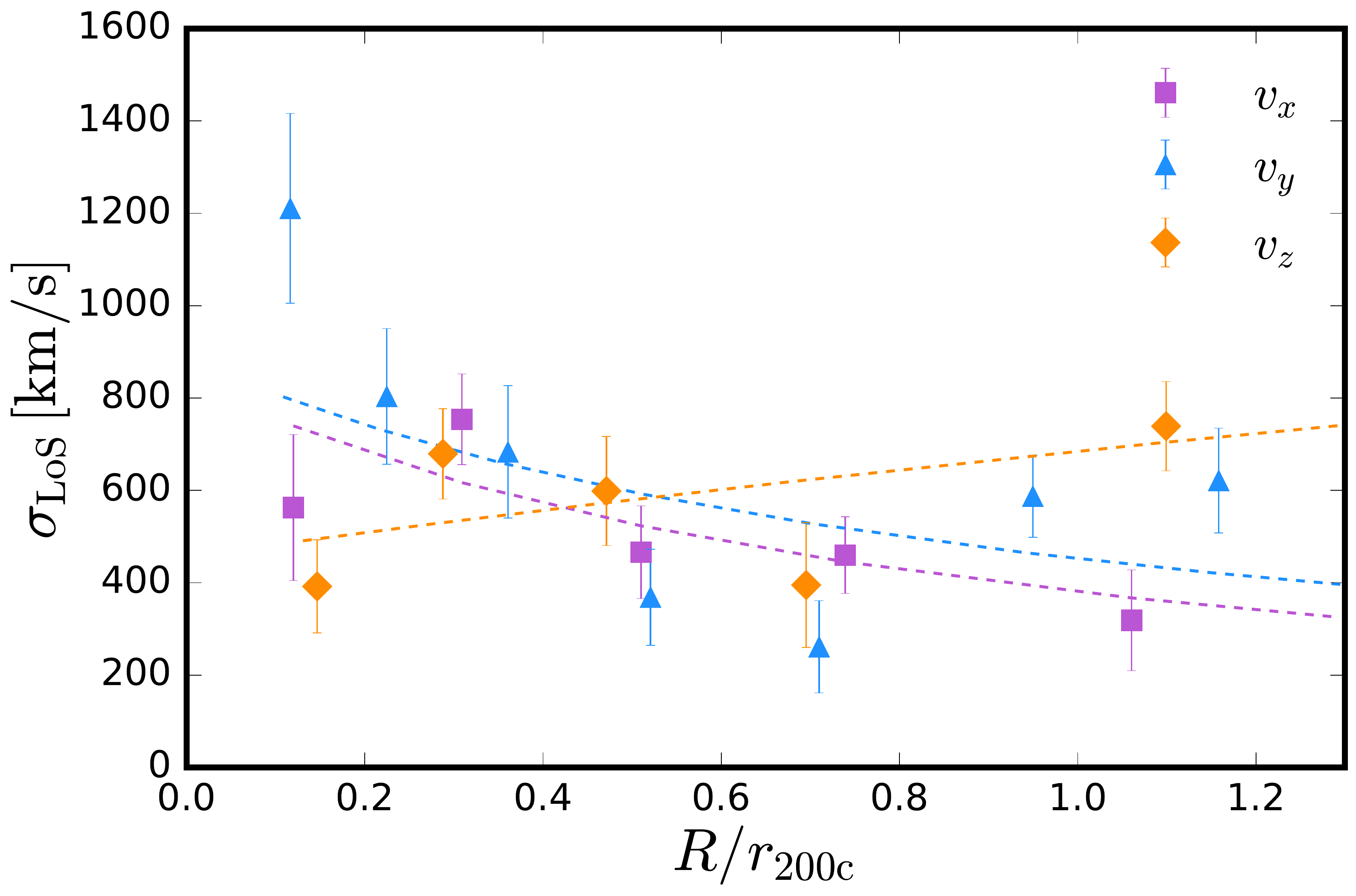}
		\caption{The LoS galaxy velocity dispersion as a function of projected radius in CE-26 (top) and CE-05 (bottom). The radial bins are scaled with respect to the true value of $r_\mathrm{200c}$. Lines show the power law fit, used to extract the gradient in the Jeans analysis. See Fig. \ref{fig:sigmaPWRAll} for all cluster profiles. The error bars denote the $1\sigma$ uncertainty of each bin, obtained via bootstrapping. While the power law model produces reasonable fits for most clusters (such as CE-26) this is not always the case, particularly with the smaller clusters, such as CE-05.}
		\label{fig:sigPWR26}
	\end{figure}

	Fig. \ref{fig:sigPWR26} shows the LoS velocity dispersion profile for three projections of CE-26 and CE-05, and the power law profile fit to them (see Fig. \ref{fig:sigmaPWRAll} for all cluster profiles). The power law is a good fit to the projected velocity dispersion profile in general, but can fail for the smaller clusters due to a limited number of galaxies, such as for CE-05 where the predicted velocity dispersion increases as a function of radius for the $\varv_z$ projection. We find a similar distribution of the exponent $p$ between the 3D and 2D cases, as seen in Fig. \ref{fig:pwrHist}.
	
	In addition to fitting parametrised models, several other methods were used to obtain the derivatives of $\nu$ and $\sigma$. Galaxy number density was found to give more consistent results compared to the mass density and is easier to obtain observationally. It is also possible to take the derivative of the $\rho \sigma^2$ profile, though this was found to give noisier results. Instead of fitting parametric models we tried smoothing the data with a Savitzky$-$Golay filter. This resulted in large and sharp changes in the measured gradients driven by noise in the profile. In summary, the NFW profile and equation (\ref{sigma_plaw}) produced the most robust fits to the data, compared to filtering or polynomials.
	
\begin{figure}
		\centering
		\includegraphics[width=0.99\linewidth]{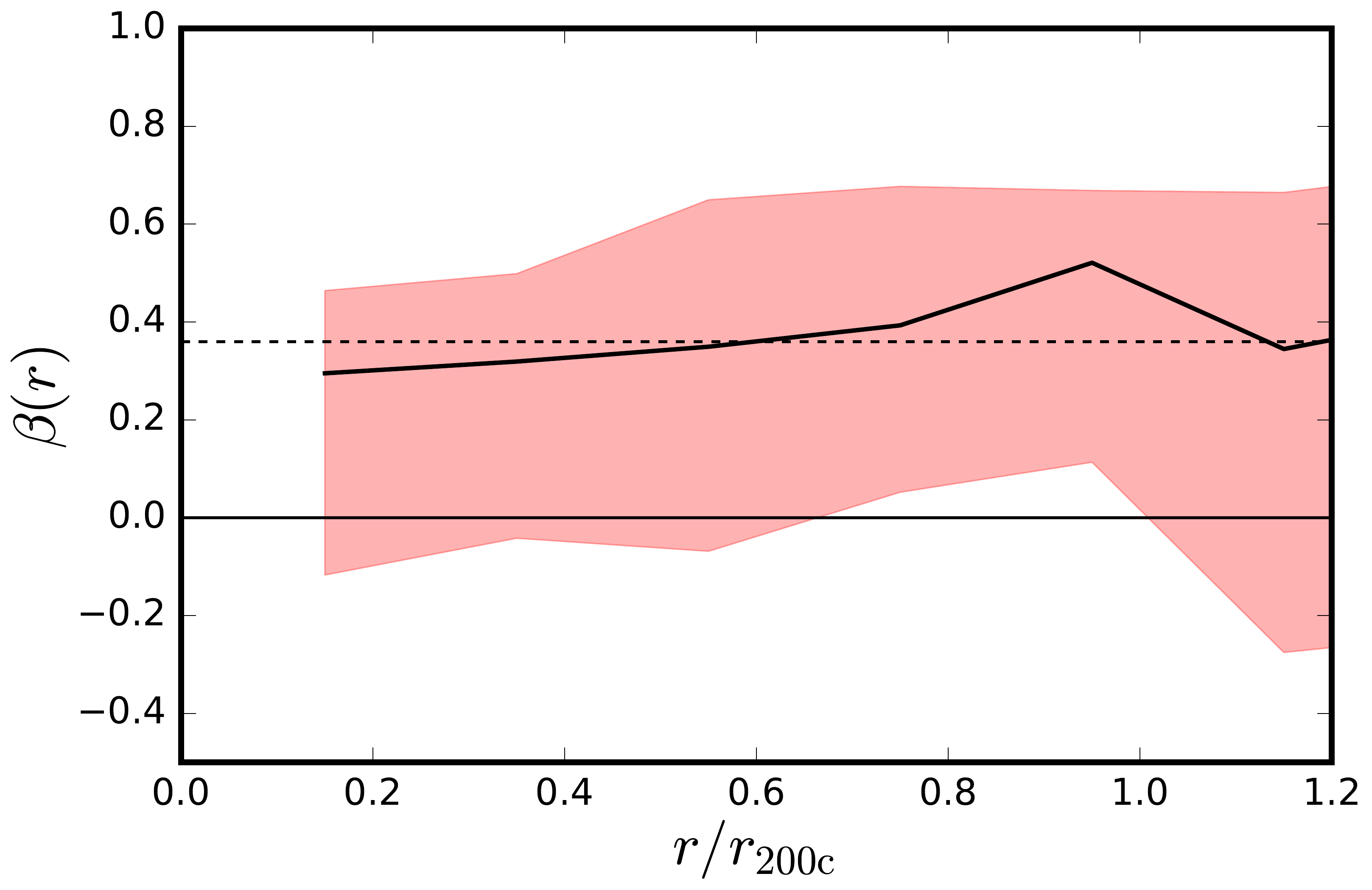}
		\caption{The median radial profile of $\beta$ for all 30 clusters. The shaded region shows the $1\sigma$ percentile spread. The horizontal dotted line is the median integrated value of $\beta=0.36$ at $r_\mathrm{200c}$. The horizontal axis denotes the true radial distance from the centre of each cluster normalised by $r_{200c}$.}
		\label{fig:betaMed}
	\end{figure}

	We cannot obtain the $\beta$ profile of the clusters in the 2D case as it requires knowledge of the three velocity components. We instead use a calibrated model with knowledge of the true profiles. We provide the true $\beta$ profiles for each of our 30 clusters in appendix \ref{app:profiles} (Fig. \ref{fig:Beta}), taking the median value of $\beta$ as a function of radius for all 30 clusters. The latter can be seen in Fig. \ref{fig:betaMed}, where it is reasonable to approximate $\beta$ as a constant. This is in line with the observations of \cite{Stark2017} who found that $\beta$ can be approximated well by a constant profile, so we adopt $\beta=0.36$ throughout this paper.
	
\begin{figure}
		\centering
		\includegraphics[width=0.99\linewidth]{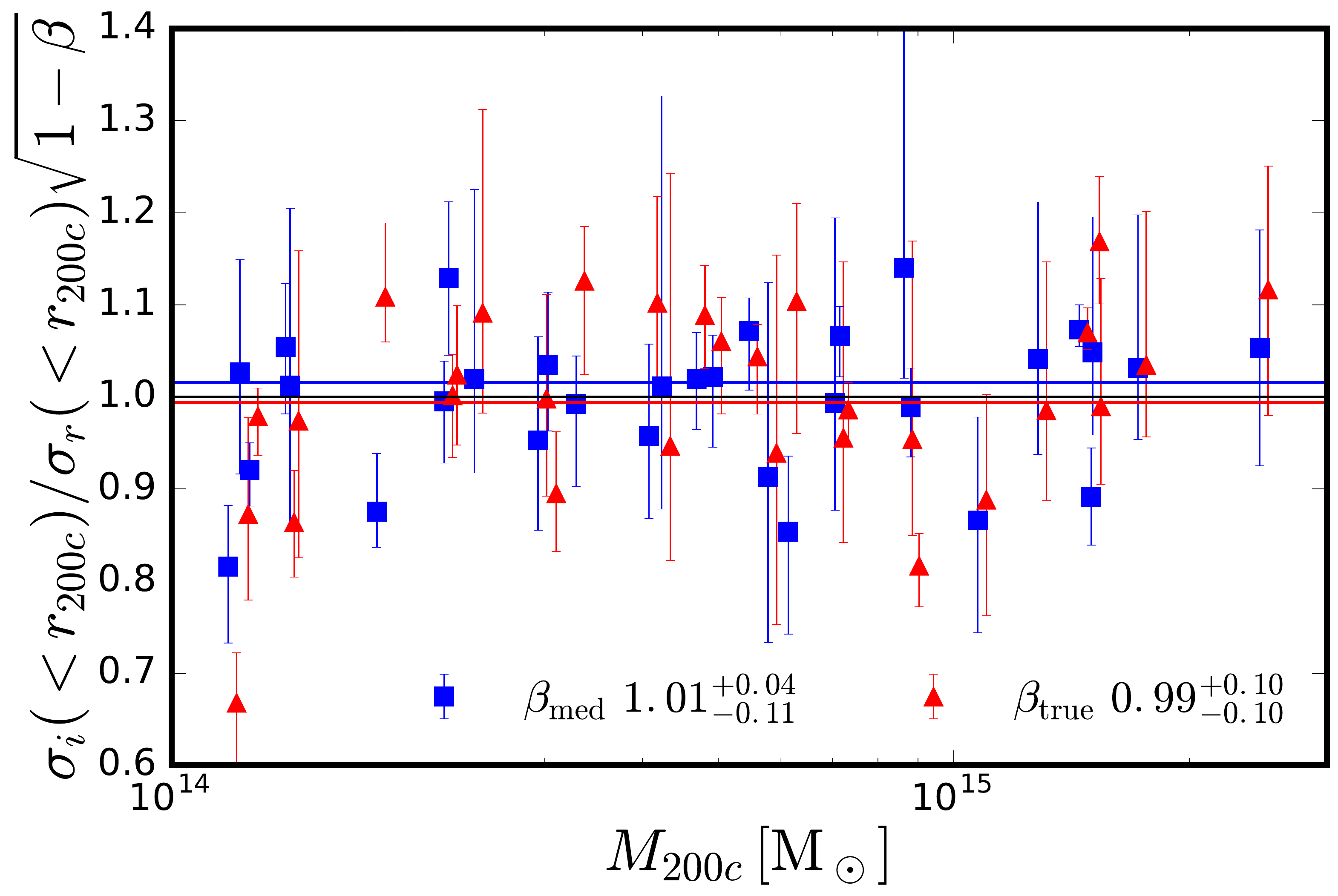}
		\caption{We show the line of sight velocity dispersion of each cluster in our sample over the radial velocity dispersion inside $r_{200c}$ corrected using equation (\ref{eqn:betaCorr}). The value of $\beta$ used for the blue squares was the median value for all clusters, 0.36, whereas the red triangles use the true value of $\beta$ for each cluster. Beside each label in the legend we show the median ratio of $\sigma_{\rm LoS}/\sigma_r$ with the $16^{\rm th}$ to $84^{\rm th}$ percentile spread.}
		\label{fig:Sigma_All_BetaCor_meVsp_med}
	\end{figure}

	In the 2D case we can only measure the line of sight velocity dispersion and so we must map $\sigma_\mathrm{LoS}$ to  $\sigma_r$ using equation (\ref{eqn:betaCorr}). We show in Fig. \ref{fig:Sigma_All_BetaCor_meVsp_med} that this assumption is valid, at least when averaged between $0-r_{200c}$. We also compare the difference between assuming a single value of $\beta=0.36$ for all clusters and using the true value of $\beta$ for each cluster. We find that when using the median value of $\beta=0.36$ the median ratio between the $\sigma_\mathrm{LoS}$ and $\sigma_r$ is $1.01^{+0.04}_{-0.11}$, whereas using the actual value of $\beta$ for each cluster gives $0.99 \pm 0.10$. In either case we recover $\sigma_r$ within the uncertainty limits.
	
	In summary, the NFW and power law fits to the density and dispersion profiles are robust when applied to the higher mass $(M_{200c}>4\times 10^{14} \; M_{\odot})$ clusters in C-EAGLE. We find that the $\beta$ profile can be assumed to be constant at $0.36$ for all clusters. This allows for an unbiased conversion between $\sigma_{\rm LoS}$ and $\sigma_r$, with a scatter of 10 per cent.
	
	\section{Cluster mass estimates} \label{res}
	We now present our main results. We first discuss estimated masses obtained using the three dynamical estimators and how they perform, both in the best possible scenario where we have full knowledge of the galaxies' 3D positions and velocities, labelled the `3D' case throughout, and for the more realistic scenario where we only have line of sight velocities and projected positions. This sample is also contaminated by interloper galaxies as our region is a cylinder of length $10r_{200c}$ centred on the cluster and projected along the line of sight. We refer to the second case as `2D'. We also test the relative performance of the techniques when we have prior knowledge of $r_{200c}$, and when $r_{200c}$ must also be estimated. We also compare masses estimated using DMO simulations of the C-EAGLE clusters, and the effects of different galaxy selection criteria. We finally consider the performance of the estimators compared to mock X-ray observations of the same clusters within the smaller aperture $r_{500c}$. Throughout this section we define the scatter, $\delta$, in dex as half the $16^{\rm th}-84^{\rm th}$ percentile range of $\log \left( M_\mathrm{est}/M_{\rm true} \right)$.
	
	\subsection{Dynamical masses} 
	\begin{figure*}
		\centering
		\includegraphics[width=0.99\linewidth]{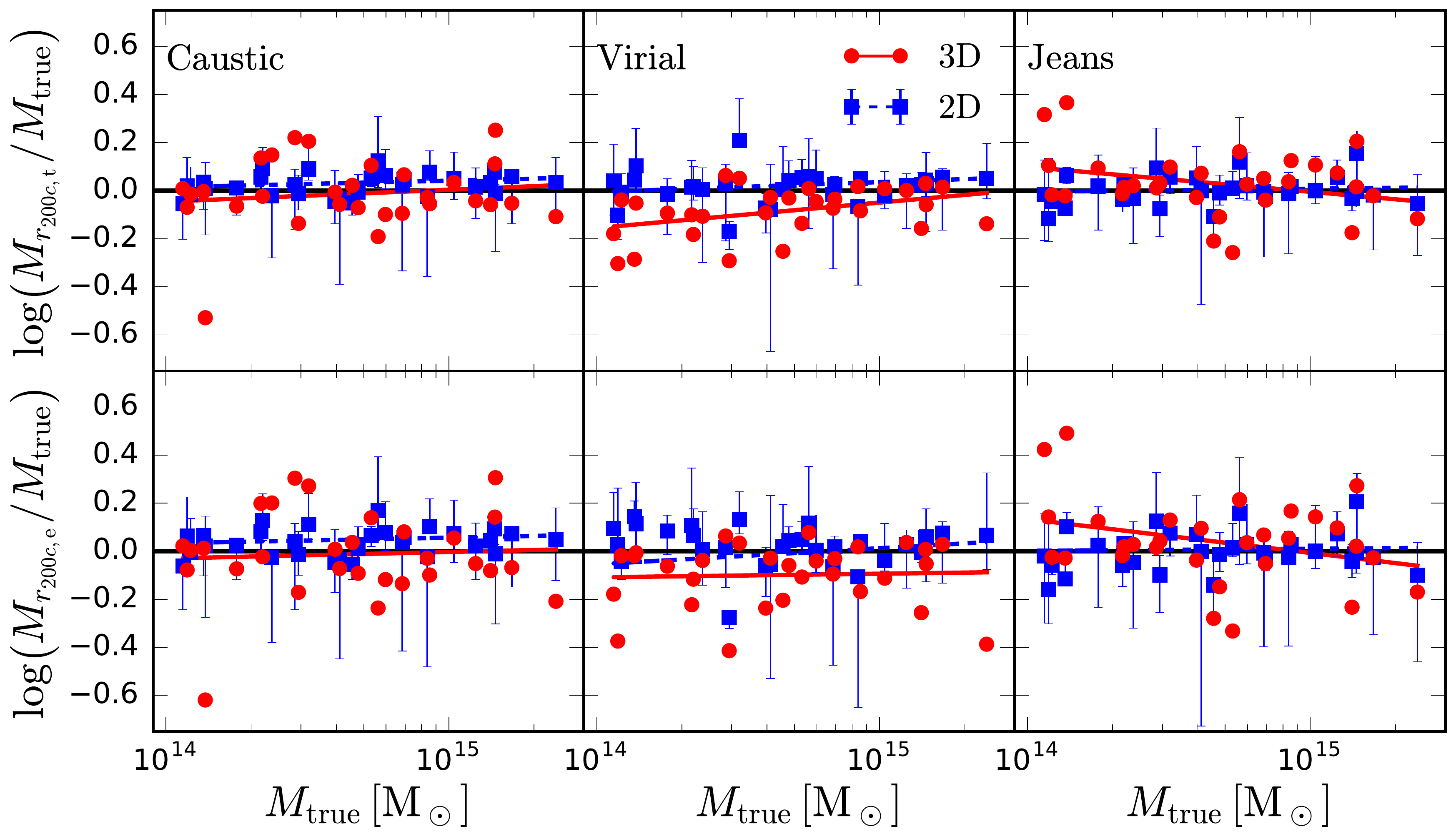}
		\caption{Top panel: the ratio between the estimated mass and the true mass in dex, within $r_{200c}$, for the 30 C-EAGLE clusters. From left to right we show the masses obtained via the caustic, virial and Jeans methods, respectively. The black solid line denotes a one-to-one relation between the estimated and true masses, while the blue squares and red circles show the values obtained using LoS and full 3D information, respectively, with a corresponding line of best fit. The error bars on the LoS points show the highest and lowest estimated masses for each cluster when projected along the different lines of sight, with the data point being the median value. The lower panel shows the same information except with the true value of $r_{200c}$ unknown and derived as part of the analysis ($r_{200c,\rm e}$).}
		\label{fig:AllMasses_tru}
	\end{figure*}
	\begin{figure}
		\centering
		\includegraphics[width=0.8\linewidth]{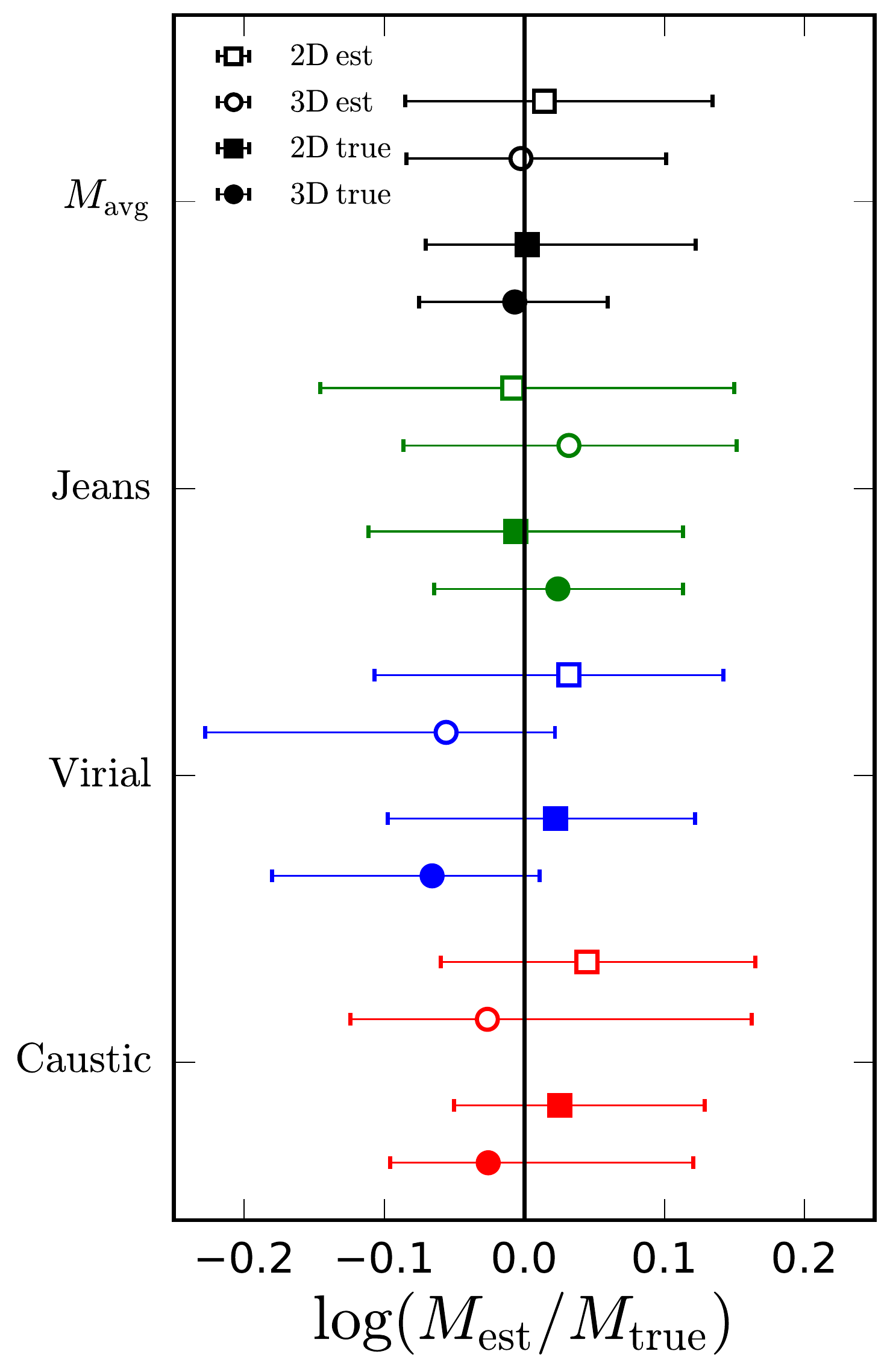}
		\caption{The median and 16/84 percentile ranges for the three different mass estimators. Circles show masses obtained using ideal 3D information and masses using 2D information are shown as squares. The solid and empty markers represent masses within the true and derived values of $r_{200c}$ respectively.}
		\label{fig:MedMassVals}
	\end{figure}
	
	\begin{table}
		\centering
		\setlength\tabcolsep{3.0pt}
		\caption{Summary of the performance of each mass estimator. Column 1 denotes the mass estimator. Columns 2 and 4 show the median ratio of the estimated cluster mass over the true mass, in dex, which we define as $\mathcal{M}=\mathrm{med} \left[ \log \left( M_\mathrm{est}/M_\mathrm{true} \right) \right]$, for the ideal 3D and realistic 2D cases respectively. The uncertainties computed via bootstrapping the sample and taking the standard deviation of the median value. Columns 3 and 5 show the scatter of the mass ratios in columns 2 and 4, with the uncertainties obtained via bootstrapping as for the median values. $r_{200c,\rm t}$ and $r_{200c,\rm e}$ denote whether the true or an estimated value of $r_{200c}$ was used in the analysis respectively. The row labelled `Average' shows the bias and scatter when the three mass estimates of each cluster are averaged together.}
		\label{M200Mes}
		\begin{tabular}{lcccc}
			\hline
			$\mathrm{Method}$   & $\mathcal{M}_\mathrm{3D}$   & $\delta_\mathrm{3D}$    & $\mathcal{M}_\mathrm{2D}$   & $\delta_\mathrm{2D}$  \\
			\hline
			$r_{200c,\rm{t}}$ \\
			$\mathrm{Caustic}$  & $-0.03\pm0.02$                        & $0.11\pm0.03$                         & $0.03\pm0.01$                         & $0.09\pm0.01$                         \\
			$\mathrm{Virial}$   & $-0.07\pm0.02$                        & $0.10\pm0.02$                         & $0.02\pm0.01$                         & $0.11\pm0.02$                         \\
			$\mathrm{Jeans}$    & $0.02\pm0.02$                         & $0.09\pm0.03$                         & $-0.01\pm0.01$                        & $0.11\pm0.01$                         \\
			$\mathrm{Average}$  & $-0.01\pm0.01$                        & $0.07\pm0.01$                         & $0.00\pm0.01$                         & $0.10\pm0.02$                         \\
			\hline
			$r_{200c,\rm{e}}$ \\
			$\mathrm{Caustic}$  & $-0.03\pm0.03$                        & $0.14\pm0.03$                         & $0.04\pm0.01$                         & $0.11\pm0.01$                         \\
			$\mathrm{Virial}$   & $-0.06\pm0.03$                        & $0.12\pm0.03$                         & $0.03\pm0.01$                         & $0.12\pm0.02$                         \\
			$\mathrm{Jeans}$    & $0.03\pm0.03$                         & $0.12\pm0.04$                         & $-0.01\pm0.01$                        & $0.15\pm0.02$                         \\
			$\mathrm{Average}$  & $-0.00\pm0.01$                        & $0.11\pm0.02$                         & $0.01\pm0.01$                         & $0.11\pm0.02$                         \\
			\hline

		\end{tabular}

	\end{table}
	
	Fig. \ref{fig:AllMasses_tru} shows the estimated values of $M_{200c}$ for all 30 C-EAGLE clusters, normalised by the true value of $M_{200c}$ for each cluster. The mass estimates from the caustic, virial and Jeans methods are shown from left to right. The red circles show the estimated mass in the ideal 3D case and the blue squares for the 2D case. The error bars denote the minimum and maximum mass estimate range for each cluster, with the square/circle denoting the median value. The solid red and dashed blue lines show the best linear fit between the estimated and true value of $M_{200c}$. Note that $r_{200c}$ is known a-priori in the upper panel, but is derived as part of the analysis ($r_{200c, \rm e}$) in the lower panel.
	
	We can see that there is significant scatter of ${\sim} 0.1-0.15 \, \mathrm{dex}$ in both the 3D and 2D cases for all methods. The scatter is quantified in Table \ref{M200Mes}. The level of scatter is similar to that seen in other simulation work, for $N_\mathrm{gal}\approx100$ (e.g. \citealt{Serra2011,Gifford2013,Gifford2013b,Mamon2013}). \cite{Old2014,Old2015} compared 25 different galaxy-based methods on a set of simulated cluster catalogues based on Halo Occupation Distribution (HOD) and Semi-Analytic Models (SAM). Their processes involved identification, interloper removal and mass estimation. They found a range of scatter between 0.18 to 1.08 dex, with the best performing phase-space methods achieving ${\sim} 0.27$ dex, which is considerably higher than our results. This is likely in part due to the additional complexity of identifying the galaxy clusters, with the lower resolution resulting in fewer galaxies. \textcolor{black}{\cite{Old2015} also define scatter as the root mean square of the logarithmic ratio between the estimated and true mass, rather than percentile spread. We find the two definitions to yield consistent values. By only selecting clusters in the \cite{Old2015} sample with at least 100 galaxies, comparable to the numbers in this work, they find dispersion based techniques to have a scatter of ${\sim}0.1$ dex, in line with our findings}. This suggests that the number of galaxies in a cluster affects the accuracy, though not enough to completely account for the difference between our results.

	Both the virial and Jeans methods suffer a major failure with CE-11 and CE-12, obtaining an estimate less than half the true mass. Similar failures happen when $r_{200c}$ is derived as well, as would be expected. However, the major failures do not all occur on the same cluster across all methods. This is due to the differing assumptions in each method; for example, the Jeans method relies on fitting multiple profiles, compared to averaged properties in the viral method. It is easy to imagine cases where the velocity dispersion profile is poorly recovered in the Jeans case but the averaged value at $r_{200c}$ is representative and vice-versa. As the three techniques all use the same initial data set it is possible to flag such major failures by cross-referencing the different masses.
	
	We find no clear evidence of a mass dependence in the mass bias for the projected samples \textcolor{black}{(similarly, we also find no dependence on richness)}. The mass trend in the fit for the 3D Jeans case is driven by the two low mass haloes which are significantly overestimated, due to the low number of galaxies present in the cluster. \textcolor{black}{As we showed in Fig. \ref{fig:MassRich}} the number of galaxies in each cluster ranges from less than 50 in the low mass end to ${\sim}$800 for the largest clusters. The virial method also suffers from a negative bias in the 3D case, with low mass clusters most strongly affected. We do caution however, that we require a significantly larger sample of clusters to say anything definitive about the mass dependency of the methods.

	Fig. \ref{fig:MedMassVals} summarises the results presented in Fig. \ref{fig:AllMasses_tru}, showing the median and $16^{\rm th}/84^{\rm th}$ percentile range for $M_\mathrm{est}/M_\mathrm{true}$, for the combinations of 2D, 3D, and true or estimated values of $r_{200c}$. The 2D cases are plotted using squares and 3D cases with circles. Whether $r_{200c}$ was provided or not is shown by filled or empty points, respectively. The colour of each point indicates the mass estimate method used, caustic (red), virial (blue) and Jeans (green). The average estimated mass for each cluster is shown by the black points and corresponds to the `Average' rows in Table \ref{M200Mes}.
	
	Providing the true value of $r_{200c}$ yields little improvement over estimating it during the analysis. The bias never changes significantly as seen in Table \ref{M200Mes}, always less than 0.07 dex. There is some evidence to support that the scatter increases when $r_{200c}$ is unknown. \cite{Mamon2013} found that their \textsc{MAMPOSS}t mass estimator also did not improve when fixing $r_{200c}$ to its true value, performing worse in some cases, arguing this to be due to halo triaxiality.
	
	We can see that, for the caustic and virial methods, including interlopers and projection effects systematically increases the estimated masses with respect to the true values, while the opposite is true for the Jeans method. In the case of the virial method we can easily identify how each of the key components, ($\sigma$, $R_{\rm H}$ and SPT) change when the data is projected and contaminated with interlopers. We found that both $R_{\rm H}$ and SPT increased in the 2D sample, though this was partially compensated by a reduction in the measured velocity dispersion. An important assumption in the Jeans 2D case is that $\beta$ is constant with radius and cluster mass. Our assumed $\beta=0.36$ is too low at $r_{200c}$, with the true median value $0.43 \pm 0.32$ at $r_{200c}$; this would contribute to a systematically lower estimate of $M_{200c}$ across all clusters. A key point to take from Fig. \ref{fig:MedMassVals} and Table \ref{M200Mes} is that taking the average of the three mass estimates for each cluster results in an unbiased estimate in all cases. 

	\subsection{Effects of baryons}
	
	\begin{figure}
		\centering
		\includegraphics[width=0.8\linewidth]{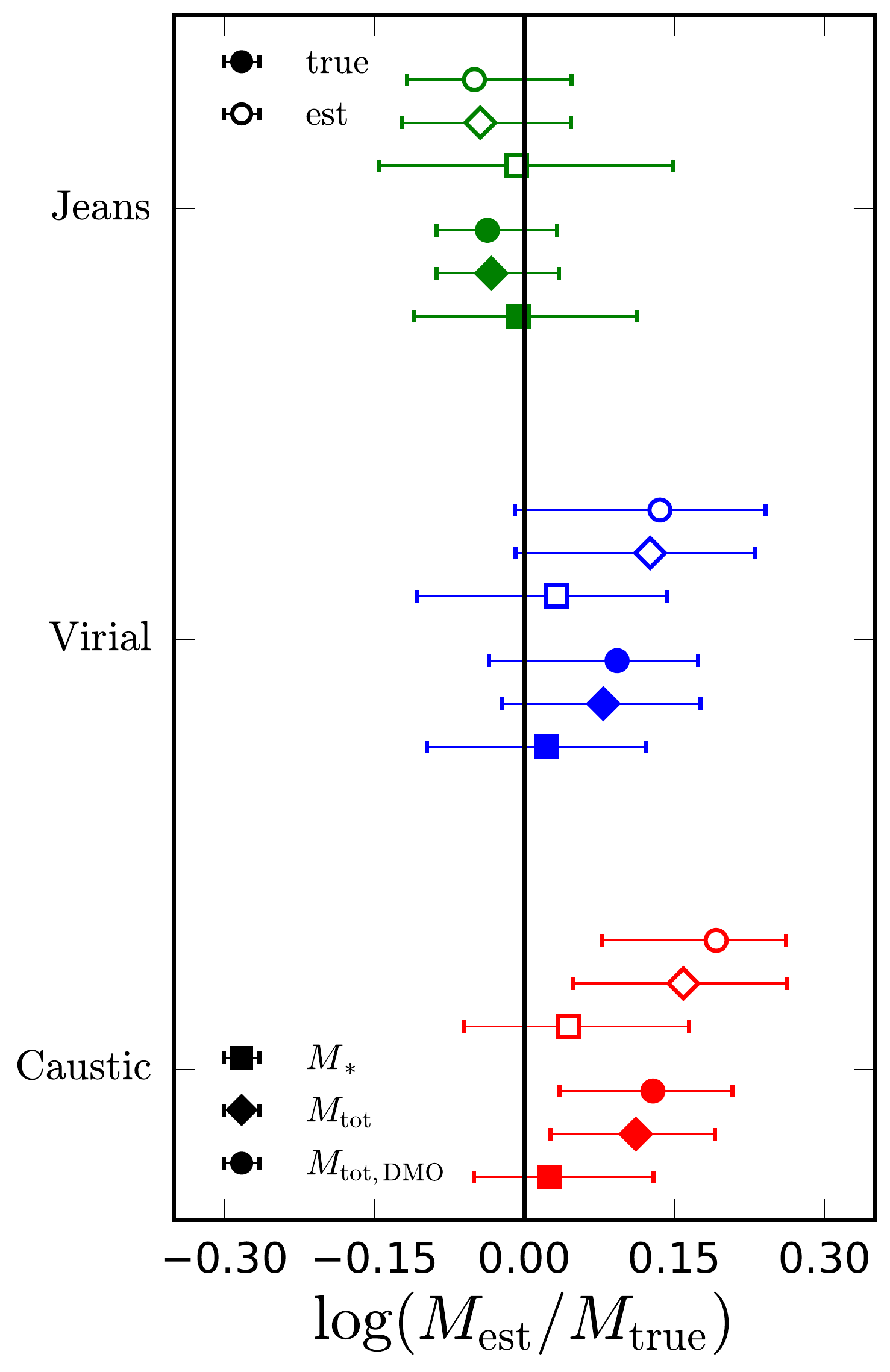}
		\caption{The median and 16/84 percentile ranges for the three different mass estimators in dex. Circles represent the DMO sample with a total mass cut of $10^{10} \, \mathrm{M_\odot}$, diamonds a total mass cut of $10^{10} \, \mathrm{M_\odot}$ in the hydro simulations and squares the stellar mass cut sample of $10^{9} \, \mathrm{M_\odot}$ used throughout the paper. The solid and empty markers represent masses within the true and derived values of $r_{200c}$ respectively.}
		\label{fig:MedMassValsBary}
	\end{figure}
	
	Due to their lower computational cost, dark matter only (DMO) simulations are often used when large volumes are required, such as forecasts for upcoming large surveys. An important question is, therefore, whether baryonic effects can be neglected in the context of dynamical mass estimates. In addition to the core hydrodynamical C-EAGLE clusters, DMO versions of all 30 clusters were also run so that the effects of the baryons can be quantified. The inclusion of baryons also allows for a more realistic selection of galaxies as it can be based upon stellar mass, rather than total mass. As shown in \cite{Armitage2018}, the selection criteria for galaxies can affect measured properties, such as the velocity dispersion, by ${\sim} 10$ per cent.
	
	Fig. \ref{fig:MedMassValsBary} shows how different selection criteria and baryonic physics affects the bias and scatter in the mass estimates. The reference sample in this paper, with a stellar mass cut of $10^9 \, \mathrm{M_\odot}$, is shown as squares, selecting by total mass in the same simulations is shown as diamonds and total mass cuts in the DMO simulations of the same clusters as circles. The total mass cut shown in Fig. \ref{fig:MedMassValsBary} was set to $10^{10} \, \mathrm{M_\odot}$ so that the number of galaxies in each sample was similar, though in practice the total mass cut sample contained approximately twice as many galaxies as the stellar mass sample. 
	
	When selecting by total mass, the bias in the caustic and virial methods increases relative to the stellar mass case. This is in line with what is seen in \cite{Armitage2018} where the velocity dispersion for stellar mass limited samples is ${\sim} 10$ per cent lower than for total mass limited samples. The estimated masses from the Jeans analysis are biased slightly lower for the total mass sample, though the difference is less than for the virial and caustic methods. The scatter is reduced by a factor of two, in this case due to the increased number of galaxies  resulting in better fits to the density and velocity dispersion profiles. (If we reduce the number of galaxies in the total mass sample to be the same as for the stellar mass sample, the scatter is increased to a similar level.) Interestingly, the scatter is similar for the caustic and virial results. This implies that the number of galaxies is not the limiting factor in the scatter. For all three methods we see little difference between the DMO and hydro simulations given the same mass cut. However, the stellar mass threshold is a better proxy for how galaxies would be selected. When galaxies are selected by their stellar mass, the bias is reduced for all three methods compared to a selection based on total mass. The difference arises due to the difference in mass loss between the stellar and DM component of a galaxy as it enters a galaxy cluster, see \cite{Armitage2018} for details.
	
	\subsection{Comparison to X-ray masses}
	\begin{figure}
		\centering
		\includegraphics[width=0.99\linewidth]{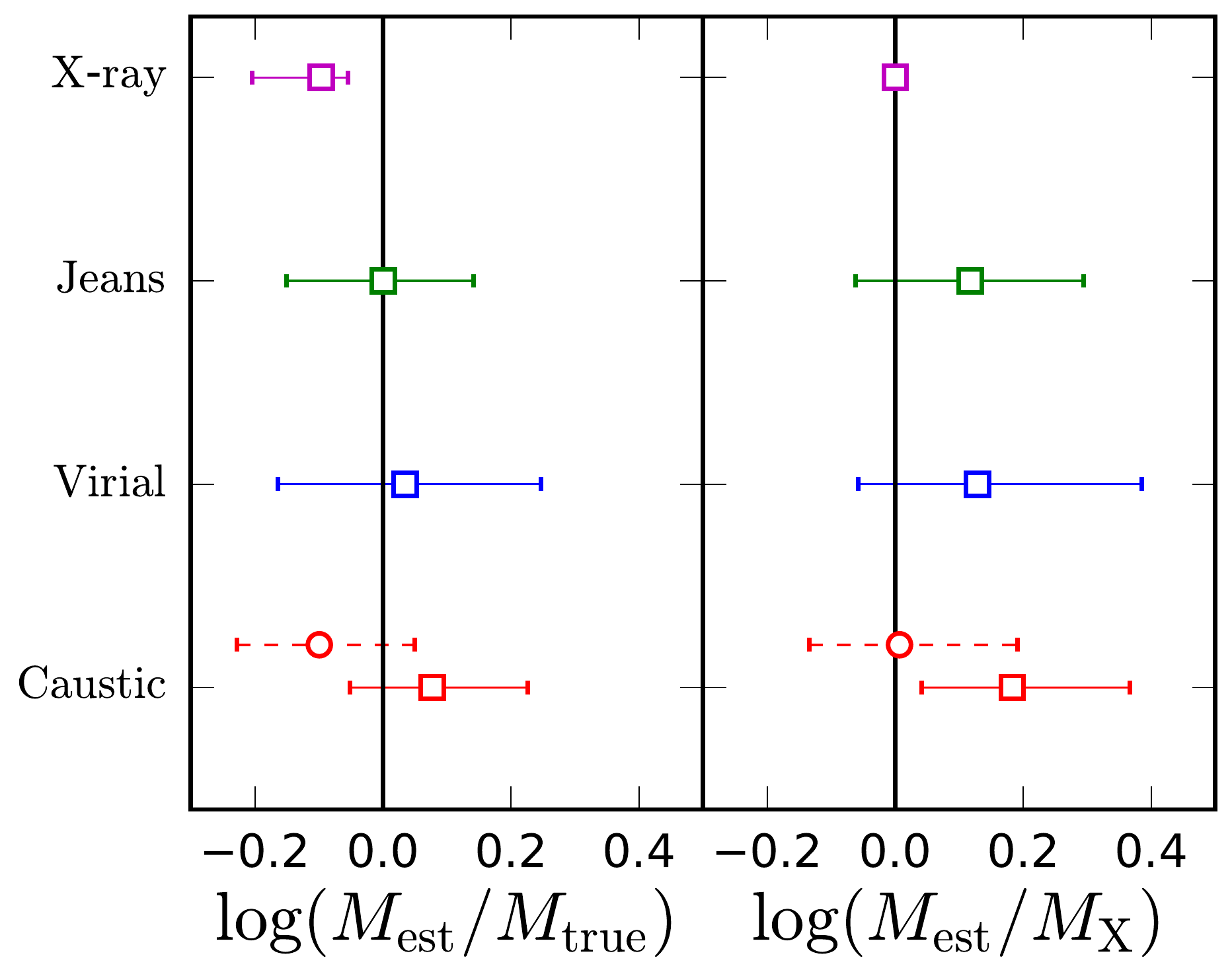}
		\caption{Left Panel: The median and 16/84 percentile ranges for the three different mass estimators for $M_\mathrm{500c}$. The spectroscopic X-ray $M_\mathrm{500c}$ is shown at the top. These masses were calculated using $r_\mathrm{500c}$ estimated individually for each method. Right panel: the median values of the three dynamical measurements normalised to the X-ray mass of each cluster. The circles with dashed error bars shows the caustic mass obtained using $\mathcal{F}_{\beta}=0.5$.}
		\label{fig:MedMassVals500}
	\end{figure}
	Mass profiles obtained from X-ray observations of clusters are an alternative to dynamical mass estimates. The two methods suffer from different systematics as they probe different components of the cluster. We now compare the estimated masses of the cluster using both X-ray and spectroscopic data. The presented hydrostatic masses are the $M_{500, \rm spec}$ values from \cite{Barnes2017b}, where the $M_{500c}$ mass has been computed by fitting X-ray derived density and temperature profiles to obtain a hydrostatic mass profile. As X-ray derived properties are typically measured inside an aperture of $r_{500c}$ we calculated $M_{500c}$ for each of our three dynamical mass estimators.
	
	Fig. \ref{fig:MedMassVals500} shows the obtained $M_{500c}$ masses using the three dynamical methods, in the 2D case without prior knowledge of $r_{500c}$, and the measured $M_{500c}$ from mock X-ray observations of the C-EAGLE clusters. For both the caustic and Jeans methods, $M_{500c}$ is obtained from the recovered mass profile, while the virial method is run using $\Delta=500$. The virial method suffers from a lack of galaxies inside this smaller aperture, increasing the scatter in the relation from $0.12 \pm 0.02$ dex to $0.21 \pm 0.02$ dex. Although the Jeans and caustic methods can use the same mass profile as for $M_{200c}$ the assumptions in each method are less justified inside this radius. The assumption that $\mathcal{F_\beta}$ is constant with radius is not a good an approximation, with the value of $\mathcal{F_\beta}=0.75$ being an overestimate at this radius. To show the effect of varying $\mathcal{F_\beta}$, Fig. \ref{fig:MedMassVals500} contains two caustic results: squares denote $\mathcal{F_\beta}=0.75$ while circles (with dashed error bars) have $\mathcal{F_\beta}=0.5$, which is at the lower end of the values used in the literature \citep{Diaferio1997,Diaferio1999}. The Jeans analysis however, is not as strongly affected by the assumptions at this range. Figs. \ref{fig:sigPWR26} \& \ref{fig:betaMed} show that the model profiles used in the Jeans analysis are valid over a wide range of radii. 
	
	The dynamical estimators suffer from slightly less bias than the X-ray measurements. We obtain broadly similar differences between X-ray and dynamical masses as \cite{Foex2017MC}. \cite{Foex2017MC} finds the Jeans method to be the least biased with respect to the hydrostatic mass, with the mean ratio $1.22 \pm 0.18$ and the virial mass differing the most, $1.51 \pm 0.26$. The median ratios for $M_{\rm est}/M_X$ are $1.3 \pm 0.1$, $1.3 \pm 0.2$ and $1.5 \pm 0.1$ for the Jeans, virial and caustic methods respectively. Given the uncertainty our results are consistent with \cite{Foex2017MC}. \cite{Maughan2016} finds caustic masses to be $20^{+13}_{-11}$ per cent larger than hydrostatic values, using $\mathcal{F_\beta}=0.5$ for their 16 clusters. In contrast, we find no significant difference between $M_{\rm est}$ and $M_X$ when $\mathcal{F_\beta}=0.5$ with a median ratio of $1.01 \pm 0.09$.

	\section{Substructure analysis} \label{sub}
	Substructure is commonly thought to be a cause of scatter in mass estimates (e.g. \citealt{Foex2017MC,Old2018}). To test whether that is the case for C-EAGLE we computed two different substructure indicators. The motivation for this is that the Jeans analysis and the virial method rely on tracing a dynamically relaxed population. However, as clusters form hierarchically, one would expect that clusters that have undergone a recent merger with another cluster or group will contain significant substructure, comprising of the remnants of the merged object. The presence of substructure is likely to increase the velocity dispersion of the cluster as the mean velocity of the substructure is unlikely to be equal to that of the host. 
	
	The methods are based upon those used in \cite{Foex2017MC}. The first considers the positive residuals after subtracting a surface density model from the cluster. The other, the Dressler-Shectman test, considers the significance of the local dynamics relative to the global averages. We first outline our implementation of these two methods and then discuss whether they are correlated with the scatter in the cluster mass estimates.
	
	\subsection{Deviation from a 2D density profile}
	Using projected positions for the galaxies, we construct a projected galaxy density map for each cluster. We represent each galaxy as a 2D Gaussian with an amplitude of unity and a dispersion of $100 \, \mathrm{kpc}$ in order to smooth the data. We take all galaxies with stellar mass greater than $10^9 \, \mathrm{M_\odot}$ along a cylinder of depth $10 r_{200c}$ and radius $2r_{200c}$, centred on the cluster centre of mass. We then fit a 2D elliptical King profile 
	\begin{equation}
	\label{KingProf}
	\mathcal{S}_{\rm King} (x,y) = \frac{\mathcal{S}_0}{1+\left(\frac{r}{r_c}\right)^2} + b  \:,
	\end{equation}
	or an NFW profile
	\begin{equation}
	\label{NFWProf}
	\mathcal{S}_{\rm NFW} (x,y) = \frac{\mathcal{S}_0}{\frac{r}{r_c}(1+\frac{r}{r_c})^2} + b  \:,
	\end{equation}
	where $\mathcal{S}_\mathrm{model} (x,y)$ is the surface density profile, $\mathcal{S}_0$ is its peak density, $r_c$ is the scale radius, and there is a uniform background term $b$. We try both King and NFW due to their differing behaviour in the central region, where NFW is cuspy and King contains a flatter core. The radial distance to a point, $r$ is formed from elliptical coordinates
	\begin{equation}
	r^2= \frac{(x \cos \phi + y \sin \phi)^2 + (y \cos \phi - x \sin \phi)^2}{(1-e)^2}  \:,
	\end{equation}
	where $e$ and $\phi$ are the ellipticity and position angle, respectively. The radial distance is scaled by the core radius $r_c$. The density profile is centred on the peak density of the cluster. This is not necessarily the same as the centre of potential, which we have taken to be the centre of the cluster throughout the analysis in the previous section. In total we fit for 5 parameters and we only constrain the parameter fitting range to physical values, i.e. all parameters must be greater than or equal to 0 and less than $2\pi$ or $1$ for $\phi$ and $e$, respectively.
	
	For each grid cell in the density map, we then subtract the model density profile leaving the residual values in each cell, $\delta_{x,y}=\mathcal{S}(x,y)-\mathcal{S}_\mathrm{mod} (x,y)$. As we want to identify the amount of substructure, i.e. an overdensity of galaxies in a given region, we calculate $\Delta$,
	
	\begin{equation}
	\Delta = \frac{\sum_{i,j} \mathrm{max}[0, \delta_{i,j}]}{\sum_{i,j} \mathcal{S}_\mathrm{model}}	\, ,
	\end{equation}
	where $\sum_{i,j}$ is a sum over all grid cells. Assuming that the number of galaxies in a cluster scales with the total mass of the cluster and that they reliably trace the underlying matter distribution, $\Delta$ should be a crude estimation of the fractional mass contained within substructures. 
	
\begin{figure}
		\centering
		\includegraphics[width=0.99\linewidth]{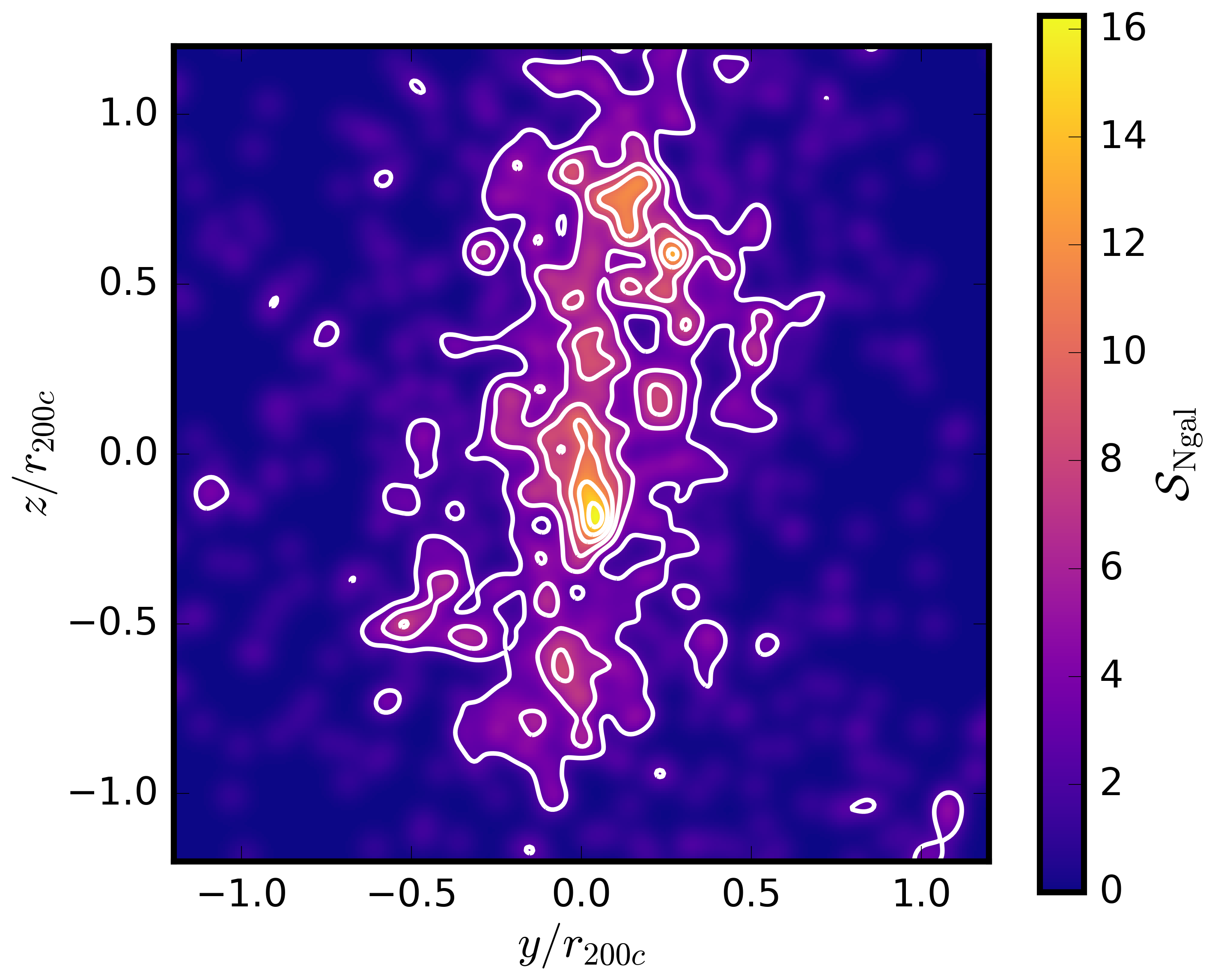}
		\includegraphics[width=0.99\linewidth]{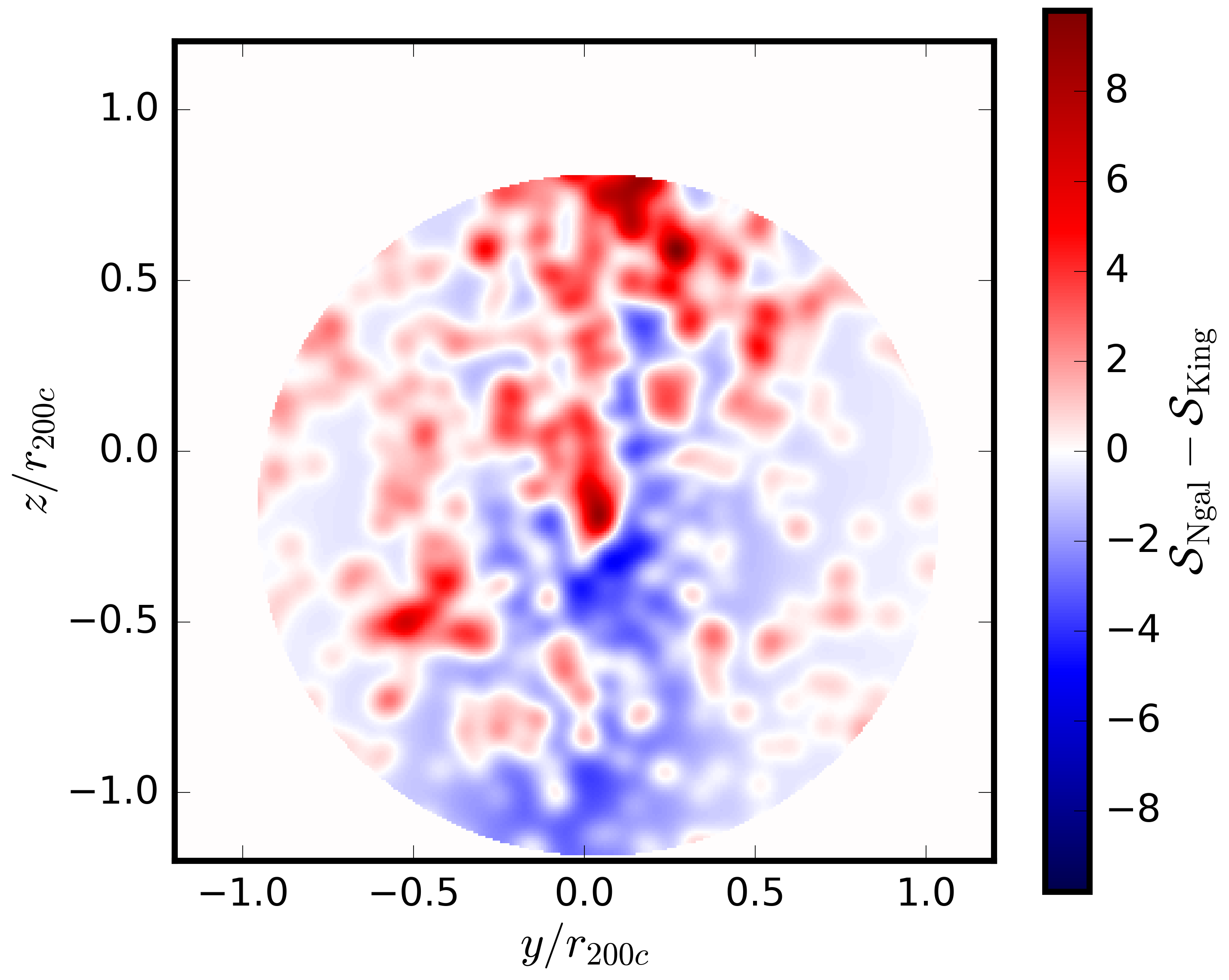}
		\caption{Top panel: example galaxy surface density map created for CE-29. The colour scale is the summed amplitude of the Gaussians representing each galaxy. Bottom panel: map of the residuals within $r_{200c}$, after subtracting an elliptical King profile. The axes are centred on the true centre of potential, whereas the profile subtraction and fitting is performed within $r_{200c}$ of the peak density, hence the offset in the vertical direction.}
		\label{fig:KingExample}
	\end{figure}

	Fig. \ref{fig:KingExample} shows one of the density maps of CE-29, the most massive C-EAGLE cluster. It is clearly an extended structure with a lot of substructure, with fit values of $e=0.72$ and $\Delta=0.28$. As can be seen from the lower panel of Fig. \ref{fig:KingExample}, where the King profile has been subtracted from the original density map, the King profile fails to reproduce the central peak. We found that the NFW profile also fails to capture the central region of clusters. However, there is little difference between predicted level of substructure when using either King or NFW, so we will present results using the King profile from now on. 
	
	\subsection{Dressler-Shectman test}
	A second metric to quantify the abundance of substructure in a cluster can be obtained by using a variant of the Dressler-Shectman test (DS: \citealt{Dressler1988}). The DS test uses both velocity and position information to identify local regions of the cluster that differ significantly from its global properties, namely the mean velocity, $\langle \varv \rangle$, and velocity dispersion, $\sigma$. \textcolor{black}{The DS test has consistently been found to be a reliable indicator of substructure \citep{Pinkney1996,Hou2009}, though \cite{White2010} found that the DS test can fail to identify substructure depending on the line-of-sight orientation. Nevertheless, in the case of groups with $N_\mathrm{gal}>20$, \cite{Hou2012} found that the DS test can be reliable when requiring a high confidence interval (95 or 99 per cent) to detect substructures (see below). For systems with $10<N_\mathrm{gal}<20$, \cite{Hou2012} conclude that the DS test can be used to obtain a lower limit on the amount of substructure.}
	
	For a given set of galaxies within a projected radius, the local mean velocity, $\varv_{\mathrm{loc}}$, and dispersion, $\sigma_{\mathrm{loc}}$, is calculated using the $n_{\rm NN}$ nearest galaxies. We follow \cite{Foex2017MC} by setting $n_{\rm NN}=\sqrt{N_{\mathrm{gal}}}$, where $N_{\mathrm{gal}}$ is the number of galaxies in the aperture. If $n_{\rm NN}<10$ then we abandon the DS test for that cluster due to an insufficient number of galaxies, which only affects the 3 of the smallest clusters. We then compute the dynamical deviation, $\gamma$, for each galaxy
	\begin{equation}
	\gamma = \sqrt{\frac{n_{\rm NN}+1}{\sigma^2} \left[ \left(\langle \varv \rangle_{\mathrm{loc}} - \langle \varv \rangle \right)^2 + \left( \langle \sigma \rangle_{\mathrm{loc}} - \langle \sigma \rangle \right)^2 \right]}  \:,
	\end{equation}
	where $\langle \varv \rangle$ and $\langle \sigma \rangle$ denote the global values of the mean velocity and velocity dispersion respectively. The next step is to bootstrap the velocities to give $10^4$ samples per galaxy. $\gamma$ is then calculated for each sample and using the $10^4 N_\mathrm{gal}$ values of $\gamma$ we compute $\gamma_\mathrm{min}$ which is defined to be the $95^\mathrm{th}$ percentile of all computed $\gamma$ values. We then define the quantity,
	\begin{equation}
	f_\mathrm{DS} \equiv \frac{N(\gamma > \gamma_\mathrm{min})}{N_\mathrm{gal}}  \:,
	\end{equation}
	which is the fraction of galaxies with $\gamma > \gamma_\mathrm{min}$, as our second substructure indicator. $f_\mathrm{DS}$ represents the fraction of galaxies whose dynamics differ significantly from the global average.
	
	\subsection{Substructure and mass estimates}
	\begin{figure}
		\centering
		\includegraphics[width=0.99\linewidth]{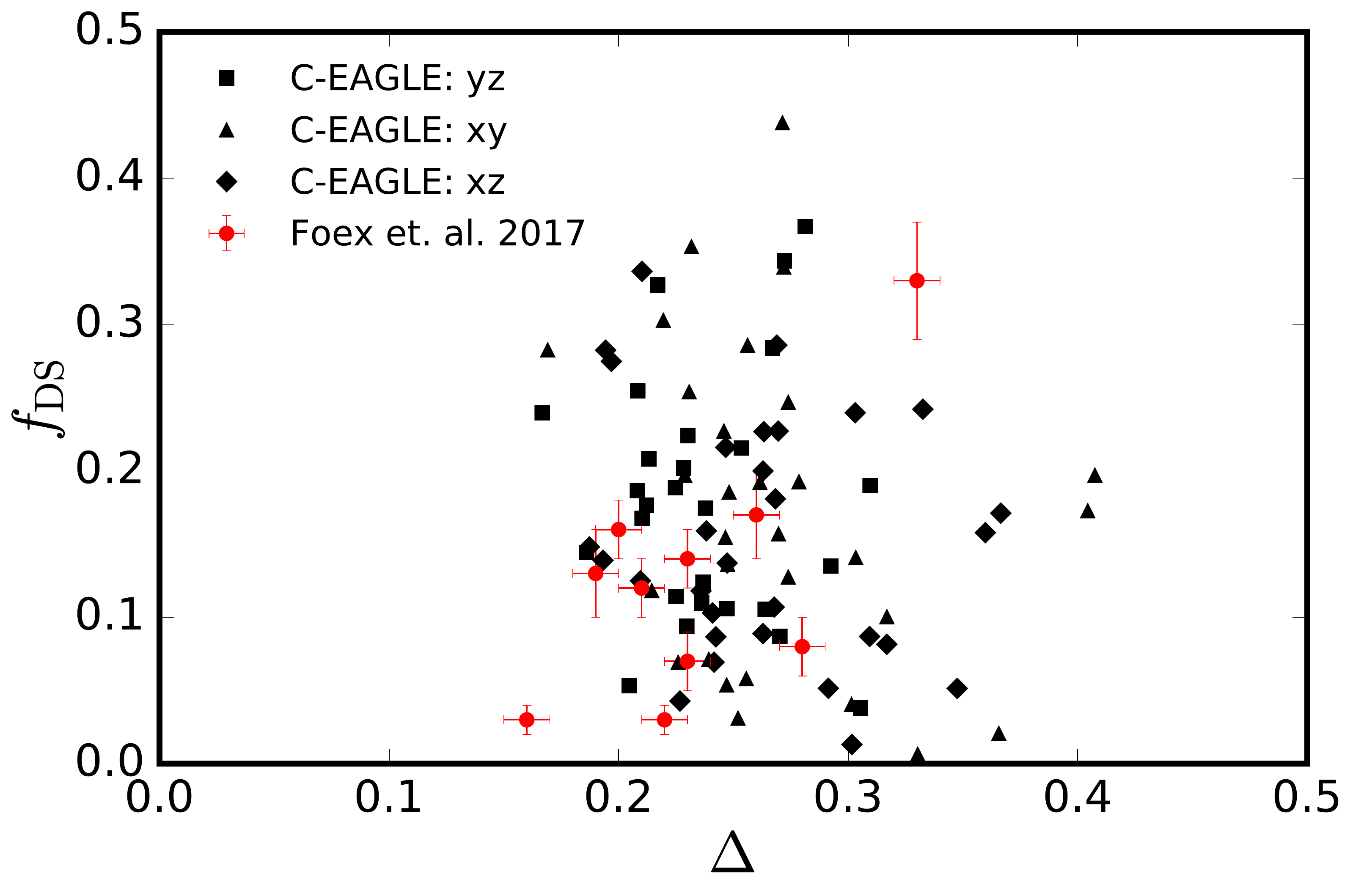}
		\caption{Comparison of the spread of values for the DS and profile tests with those found in \protect\cite{Foex2017MC}. The red circles are the substructure values found in \protect\cite{Foex2017MC}, and the black squares, triangles and diamonds are the values found for the 3 different projections of C-EAGLE data.}
		\label{fig:SubstructFoex}
	\end{figure}
	\begin{figure}
		\centering
		\includegraphics[width=0.99\linewidth]{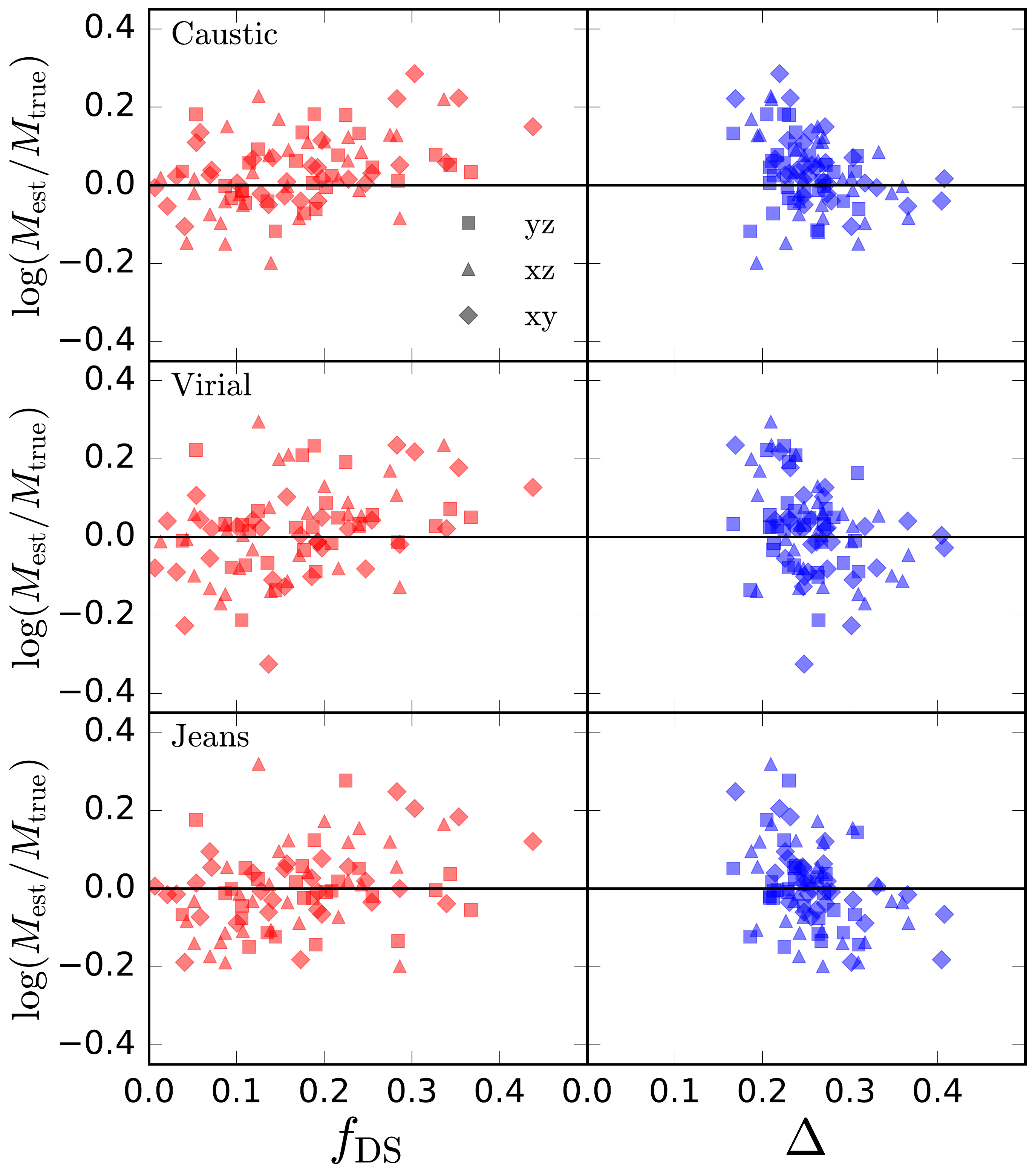}
		\caption{A comparison of two substructure indicators, the DS test and the summed differences from a elliptical King profile, $\Delta$ and how they correlate with mass bias. We plot the fractional difference between the observed mass for a given method against the amount of substructure. These are all projected quantities using the masses calculated inside the true value of $r_{200c}$ for each cluster. Going from top to bottom, the obtained masses are from the caustic, virial and Jeans methods, respectively.}
		\label{fig:Substructure_Bias}
	\end{figure}
	
	\begin{table}
		\centering
		\caption{The Pearson coefficients between the mass bias of each method and the amount of substructure with respect to the two metrics, $f_\mathrm{DS}$ and $\Delta$. The errors are obtained through $10^4$ bootstrap resampling of the clusters.}
		\begin{tabular}{lcccc}
			\hline
			$\mathrm{Method}$   & $f_{\mathrm{DS}}$   & $\Delta_{\rm King}$   \\
			\hline
			$\mathrm{Caustic}$  & $0.411\pm0.09$      & $-0.377\pm0.08$       \\
			$\mathrm{Virial}$   & $0.309\pm0.09$      & $-0.352\pm0.08$       \\
			$\mathrm{Jeans}$    & $0.321\pm0.10$      & $-0.415\pm0.07$       \\
			\hline
		\end{tabular}
		\label{substructTable}
	\end{table}
	
	\begin{figure}
		\centering
		\includegraphics[width=0.99\linewidth]{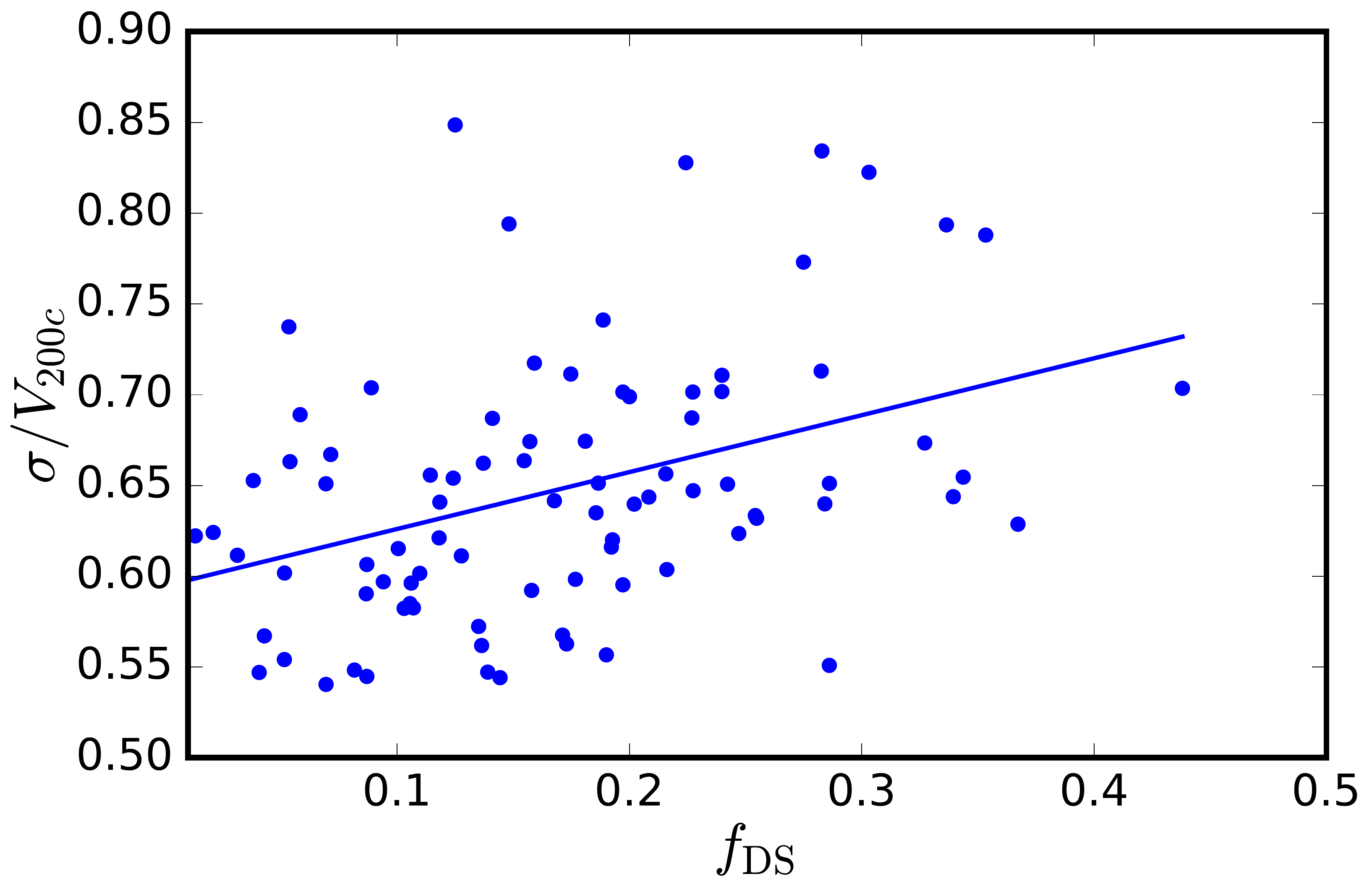}
		\caption{The line of sight velocity dispersion of each cluster normalised with respect to $V_{200c}=\sqrt{GM_{200c}/r_{200c}}$, against $f_{\rm DS}$, where we use the true values of $M_{200c}$ and $r_{200c}$. Each cluster has been projected along three orthogonal axes and the line of best fit has been calculated using all 90 projections.}
		\label{fig:sigTotfDS}
	\end{figure}
	
	As shown in Fig. \ref{fig:SubstructFoex}, the spread of the two substructure indicators, $\Delta$ and $f_{\rm DS}$, is somewhat comparable to that seen in \cite{Foex2017MC}, though we find many more extreme values of both $\Delta$ and $f_\mathrm{DS}$ with our 90 projections compared to the 10 clusters in \cite{Foex2017MC}. We show the C-EAGLE values as black points with the three different marker symbols corresponding to different projections of the same clusters and the observed \cite{Foex2017MC} as red circles; the error bars are standard deviations as obtained in the paper. We compute a two-sample KS statistic for both $f_{\rm DS}$ and $\Delta$, finding a value of 0.40 and 0.42, respectively, between the C-EAGLE and \cite{Foex2017MC} clusters, with p values of 0.08 and 0.06. The critical value to reject the hypothesis that both datasets are drawn from the same distribution with 95 per cent confidence is $0.45$. As the KS statistic for both indicators is lower than the critical value we cannot say that our substructure indicators are inconsistent with those of \cite{Foex2017MC}.

	The two substructure indicators also (weakly) correlate with mass bias, as seen in Fig. \ref{fig:Substructure_Bias} and Table \ref{substructTable}. However, the trends are in the opposite direction from each other: a greater $f_\mathrm{DS}$ implies a larger over-prediction of cluster mass, whereas lower values of $\Delta$ correspond to over-predictions of mass. 
	
	The DS test is easier to explain, as this test primarily probes velocity substructure. This relates to the velocity dispersion; if a cluster contains significant velocity substructure then the velocity dispersion would increase. Fig. \ref{fig:sigTotfDS} shows how the velocity dispersion of a cluster increases as a function of $f_{\rm DS}$. This is the primary cause of the correlation seen in Fig. \ref{fig:Substructure_Bias}. \cite{Old2018} find a similar relationship, where clusters with a high value of $f_{\rm DS}$ are systematically biased high relative to clusters with low $f_{\rm DS}$ by ${\sim} 10$ per cent.
	
	The negative correlation of the $\Delta$ statistic and mass bias is more complex. We considered both the King and projected NFW profiles to see if using a cuspy or cored profile affects the results. We found little difference between the two profiles, both showing the negative correlation. There is a weak positive correlation between the ratio of the scale radius of the galaxy number density profile and the scale radius obtained from using all particles in a cluster and $\Delta$. In the Jeans analysis an overestimated $r_{s}$ would result in a lower mass estimation of the cluster.
	
	Above a certain value of $\Delta$ the mass bias is unchanging, particularly for the caustic method. The Jeans analysis is the one that is most affected by high values of $\Delta$, though this is likely due to the dependence on the galaxy number density profile, as previously mentioned. The other two methods are likely biased high at low $\Delta$ due to the greater effect a few spurious galaxies will have on the dynamical analysis, which would tend to increase the velocity dispersion.
	
	In summary, we find that there is a weak and very noisy correlation between the presence of substructure and mass bias. We would require a larger sample of clusters in order to draw more meaningful conclusions. We should also note that as we are limited by the volume of the high resolution region in the simulation, we have not been able to fully replicate the presence of interlopers and their removal. However, as we mentioned earlier, we find that our results change little when we include interlopers out to $10r_{200c}$ for the 13 \textit{Hydrangea} clusters \citep{Bahe2017} in our sample.
	
	\section{Conclusions} \label{conc}
	In this paper, we have used the C-EAGLE suite of 30 galaxy clusters \textcolor{black}{with median mass, $M_{200c}=10^{14.7} \, \rm M_\odot$,} to quantify the accuracy of three dynamical cluster mass estimators. The simulated clusters are amongst the highest resolution clusters to date ($\sim$1 kpc force resolution at $z\approx0$), and use the EAGLE subgrid physics model, which has been calibrated to reproduce the stellar mass functions, sizes, and black hole masses of field galaxies \citep{Schaye2015,Crain2015}. Previous work has also shown that the C-EAGLE clusters reproduce many properties of cluster galaxies \citep{Bahe2017} and the ICM \citep{Barnes2017b}.
	
	The three dynamical estimators tested in this paper are the caustic, Jeans and virial methods. Our galaxy tracer population was selected by a stellar mass limit of $10^9 \, \mathrm{M_\odot}$, chosen as this represents a reasonable limit with respect to upcoming surveys, such as \textit{Euclid} \citep{Laureijs2011}\textcolor{black}{, for a `golden' sample, with a median of ${\sim}180$ galaxies in a cluster}. We performed the analysis in two different scenarios, the ideal case, with full knowledge of galaxy positions and velocities to give an upper performance limit, and a more realistic case of LoS velocities, projected positions and interloper contamination out to $5 r_{200c}$. We also compared the mass obtained with and without prior knowledge of $r_{200c}$. \textcolor{black}{We should emphasize that we did not employ an interloper removal scheme in this work. Due to the high mass of the C-EAGLE clusters and limited volume, the impact of interloper galaxies was found to be minimal.}
	
	This paper also discusses the effect of substructure and how it correlates with mass bias. We employ two substructure metrics as described in \cite{Foex2017MC}. The $f_{\rm DS}$ and $\Delta$ metrics quantify substructure primarily using velocity and spatial information, respectively.
	
	Our main findings can be summarised as follows:
	\vspace{-0.1cm}
	\begin{enumerate}
		
		
		\item The LoS velocity dispersion profile is well modelled by $\sigma_0(1+r)^p$ (Fig. \ref{fig:sigPWR26}). The radial velocity dispersion can be effectively recovered via $\sigma_r=\sigma_{\rm LoS}/\sqrt{(1-\beta)}$ (Fig. \ref{fig:Sigma_All_BetaCor_meVsp_med}) to within one percent accuracy on average.
		
		\item On average, the velocity anisotropy profile, $\beta(r)$, can be assumed to be constant for a given cluster (Figs. \ref{fig:betaMed} \& \ref{fig:Beta}). The median value across the 30 clusters was $0.36$. While this is not a good approximation for a few clusters, such as CE-27, this simplifies many observations that rely on knowing $\beta$, which is difficult to measure observationally.
		
		\item The three mass estimators perform similarly with and without prior knowledge of $r_{200c}$ (Table \ref{M200Mes}, Fig. \ref{fig:AllMasses_tru}). When the results are averaged, the mass estimators are unbiased, but the scatter is significant, between 30 and 35 per cent in the projected case when $r_{200c}$ is unknown. It should be noted that even in the ideal scenario, the scatter never reduces below ${\sim} 20$ per cent. \textcolor{black}{We also do not account for cluster surveys being incomplete, using all galaxies with stellar mass $> 10^9 \, \rm M_\odot$ in our sample. As such our quoted scatter values represent the best case scenario in that regard.}
		
		\item We find no significant difference between the masses obtained using DMO and hydro simulations, when using identical (total) mass cuts (Fig. \ref{fig:MedMassValsBary}). Selecting galaxies by their stellar mass reduces the bias for all three methods, a selection that is not possible to mimic with DMO simulations.
		
		\item Comparing the X-ray $M_{500c}$ masses with the dynamical estimates for $M_{500c}$ (Fig. \ref{fig:MedMassVals500}), we find a large scatter in the dynamical mass estimates relative to the X-ray. However, the mean bias is larger for the X-ray as $M_X/M_{\rm true}{\sim} 0.8$, but with large scatter. Scatter between the X-ray and dynamical masses is ${\sim} 60$ per cent. The bias observed in the caustic method depends sensitively on the value of $\mathcal{F}_\beta$ chosen. All three dynamical methods are limited in this comparison by the lack of galaxies within $r_{500c}$.
		
		\item Two substructure identification methods, based on \cite{Foex2017MC}, show a weak correlation with mass bias (Fig. \ref{fig:Substructure_Bias}). The DS test is positively correlated, with the main driver being the overestimation of the velocity dispersion due to additional velocity substructure (Fig. \ref{fig:sigTotfDS}). The surface density residuals show a more complex dependence and require further study to determine the root cause.

	\end{enumerate}
	
	In conclusion, our simulations suggest that dynamical mass estimation techniques are a competitive alternative to X-ray hydrostatic and weak lensing methods when high quality spectroscopic data are available (our results were derived for a stellar mass limit of $10^{9} \, \rm M_\odot$). In particular, by combining the three different dynamical mass estimators, it is possible to obtain an unbiased estimate of the cluster mass on average. However, individual clusters can have masses that are biased by around 25 per cent within $r_{200c}$. Care must also be taken with choosing the values of $\mathcal{F_\beta}$ and $\beta$. In future work, we plan to assess whether the scatter in cluster mass estimates can be reduced \textcolor{black}{through the application of machine learning techniques, building on the work of \cite{Ntampaka2015,Ntampaka2016}.}
	
	\section*{Acknowledgements}
	This work used the DiRAC Data Centric system at Durham University, operated by the Institute for Computational Cosmology on behalf of the STFC DiRAC HPC Facility (www.dirac.ac.uk). This equipment was funded by BIS National E-infrastructure capital grant ST/K00042X/1, STFC capital grants ST/H008519/1 and ST/K00087X/1, STFC DiRAC Operations grant ST/K003267/1 and Durham University. DiRAC is part of the National E-Infrastructure. The Hydrangea simulations were in part performed on the German federal maximum performance computer ``HazelHen'' at the maximum performance computing centre Stuttgart (HLRS), under project GCS-HYDA / ID 44067 financed through the large-scale project ``Hydrangea'' of the Gauss Center for Supercomputing. Further simulations were performed at the Max Planck Computing and Data Facility in Garching, Germany. We also gratefully acknowledge PRACE for awarding the EAGLE project access to the Curie facility based in France at Tr\'es Grand Centre de Calcul. YB acknowledges funding from the European Union's Horizon 2020 research and innovation programme under the Marie Sk\l{}odowska-Curie grant agreement number 747645 (ClusterGal) and the Netherlands Organisation for Scientific Research (NWO) through VENI grant 016.183.011. DJB and STK acknowledge support from STFC through grant ST/L000768/1. TJA is supported by an STFC studentship.
	
	
	\bibliographystyle{mnras}
	\bibliography{BibMain}
	
	
	
	\appendix
	\section{Cluster profiles} \label{app:profiles}
\begin{figure*}
		\centering
		\includegraphics[width=0.99\linewidth]{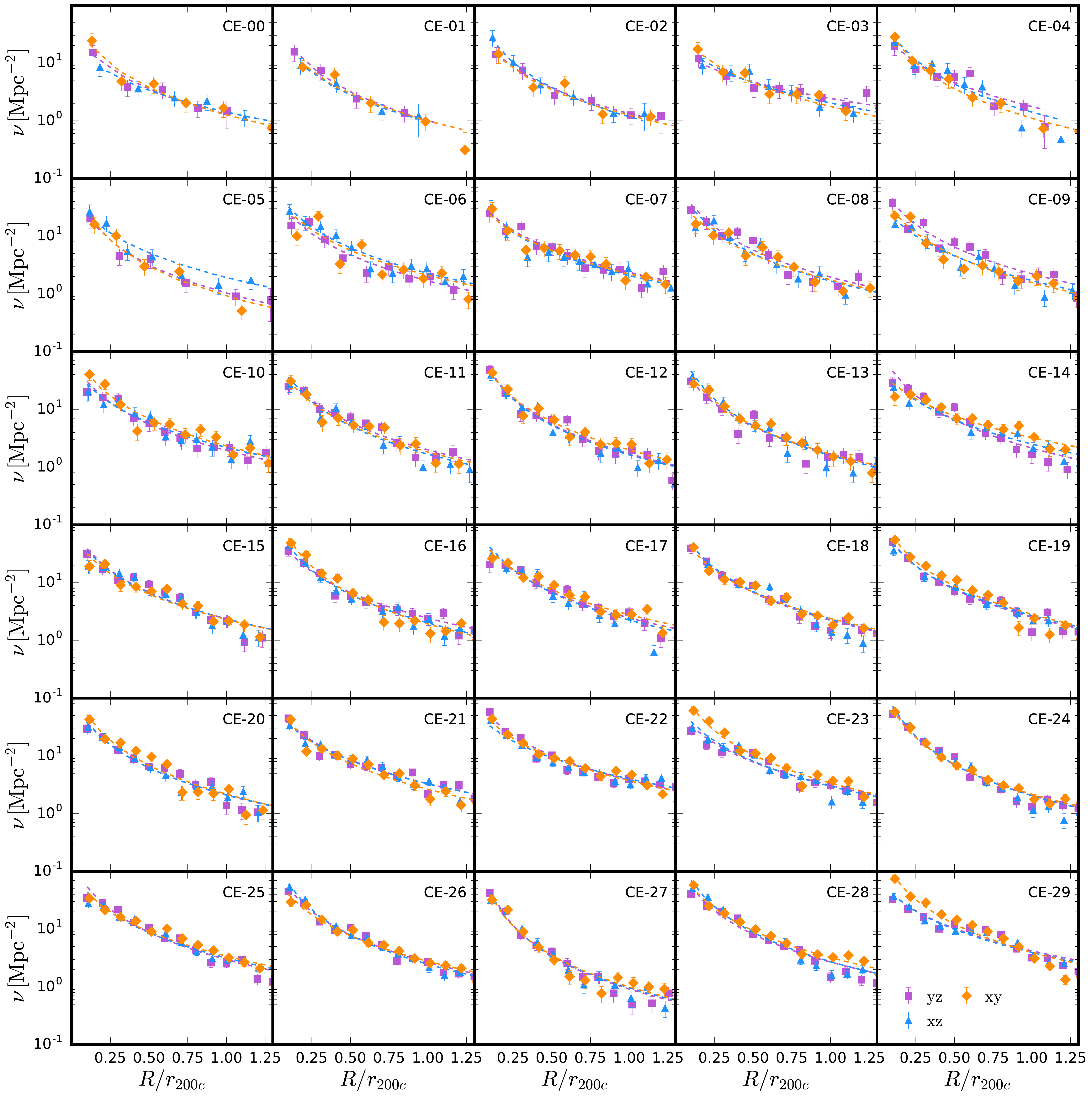}
		\caption{The projected surface density of galaxies as a function of radius for all C-EAGLE clusters. The projected radial bins are scaled with respect to the true value of $r_\mathrm{200c}$. The lines show the recovered profile, used to extract the gradient in the Jeans analysis.}
		\label{fig:SurfDens}
	\end{figure*}

	Here, we show all 30 cluster profiles used in the Jeans analysis. Fig. \ref{fig:SurfDens} shows the projected surface density profiles of the C-EAGLE clusters and the measured fits assuming a projected NFW profile. The squares, triangles and diamonds show the data for three orthogonal projections of the cluster. These profiles are contaminated by interloper galaxies within $\pm5R_{200c}$ of the cluster centre.
	
\begin{figure*}
		\centering
		\includegraphics[width=0.99\linewidth]{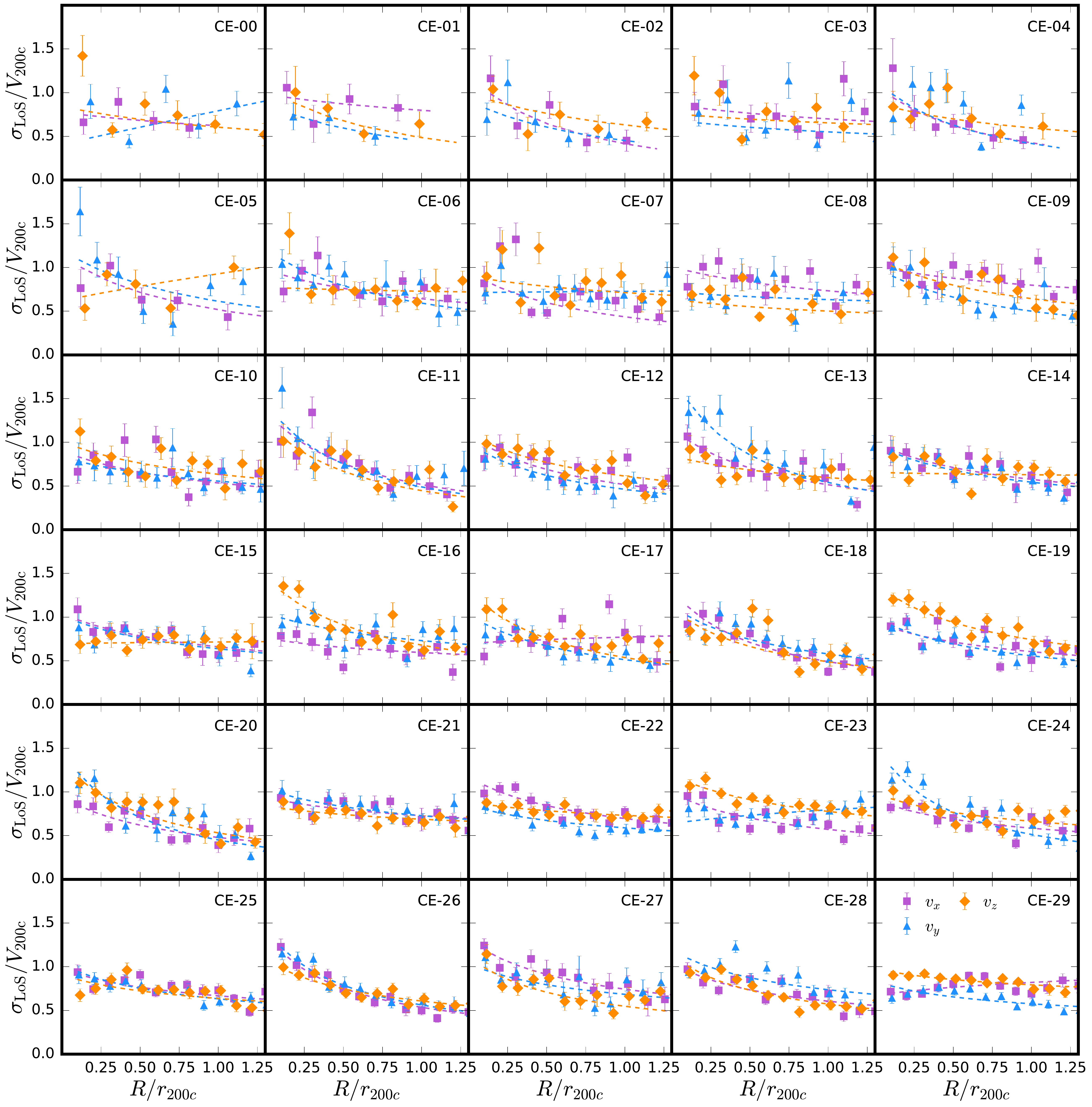}
		\caption{The projected velocity dispersion of galaxies as a function of radius for all C-EAGLE clusters. The projected radial bins are scaled with respect to the true value of $r_\mathrm{200c}$ and the velocity dispersion scaled with $V_\mathrm{200c}=\sqrt{GM_\mathrm{200c}/r_\mathrm{200c}}$. The lines show the power law fit, used to extract the gradient in the Jeans analysis.}
		\label{fig:sigmaPWRAll}
	\end{figure*}

	Fig. \ref{fig:sigmaPWRAll} shows the projected velocity dispersion profiles, for all 30 C-EAGLE clusters, in the same manner as for Fig. \ref{fig:sigPWR26}. The fitted model is equation (\ref{sigma_plaw}) and the scale is normalised to $V_{200c}$ for each cluster. There are several instances where the data is too noisy to obtain a reliable fit to the dispersion profile, particularly for the lower mass clusters, where in some cases the velocity dispersion profile increases with projected radial distance. This is driven by an intrinsic lack of galaxies in these clusters.
	
\begin{figure*}
		\centering
		\includegraphics[width=0.99\linewidth]{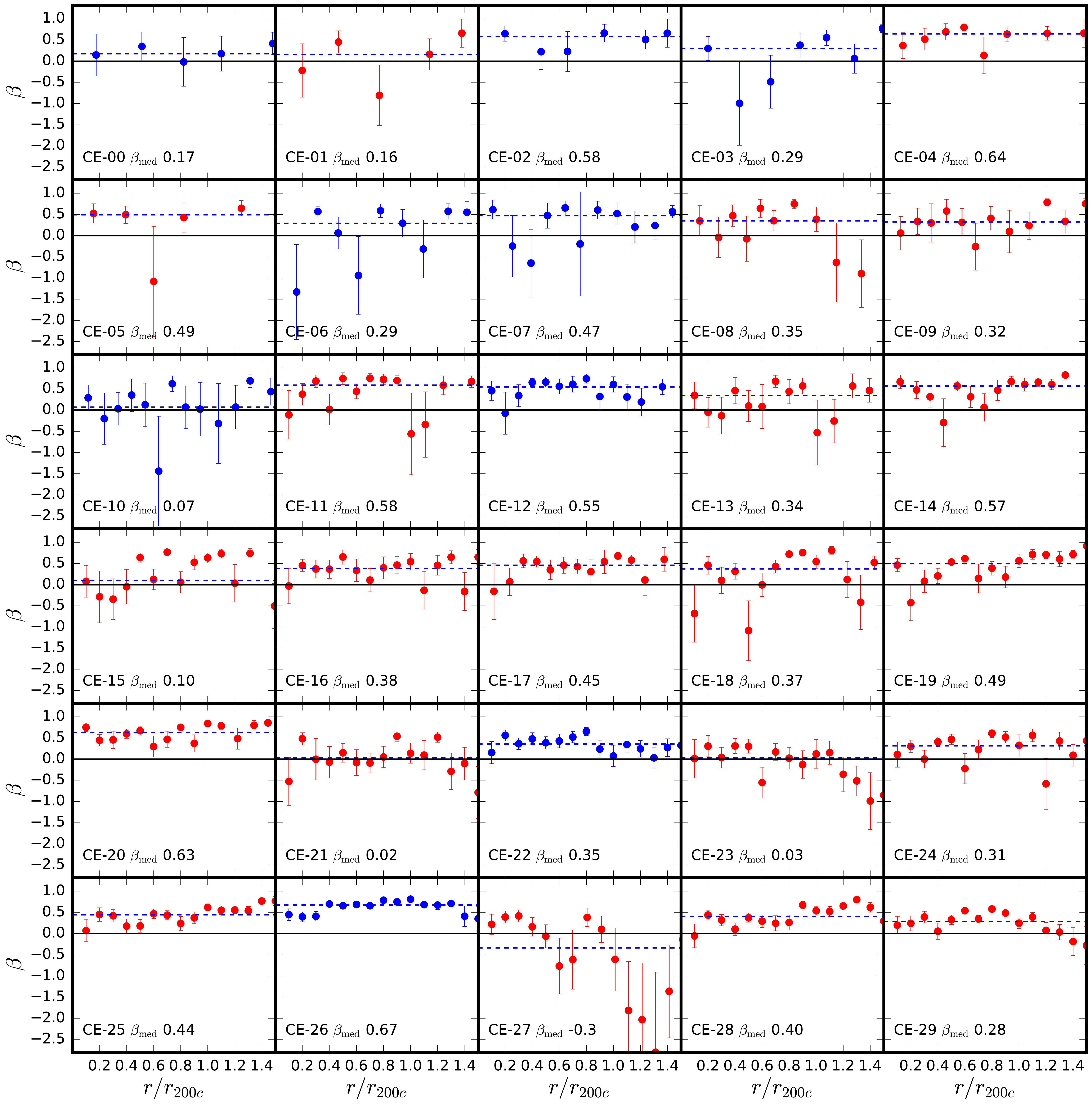}
		\caption{True velocity anisotropy profiles for all 30 galaxy clusters. The horizontal lines represent the weighted mean values. Blue represents dynamically relaxed clusters, using the ratio of kinetic to thermal energy criteria in \protect\cite{Barnes2017b}.}
		\label{fig:Beta}
	\end{figure*}

	Fig. \ref{fig:Beta} is the velocity anisotropy profile measured using the galaxies inside the true $r_{200c}$. Blue clusters are dynamically relaxed according to the ratio of thermal to kinetic energy as defined by \cite{Barnes2017b}, whereas red denotes unrelaxed. We find that the $\beta$ profile is largely flat in most clusters out to ${\sim} r_{200c}$. This is in contrast to what is observed in the DMO simulations, where the $\beta$ profile tends to rise as a function of radius.
	
	\section{Distribution of Jeans fit Parameters}
\begin{figure}
		\centering
		\includegraphics[width=0.99\linewidth]{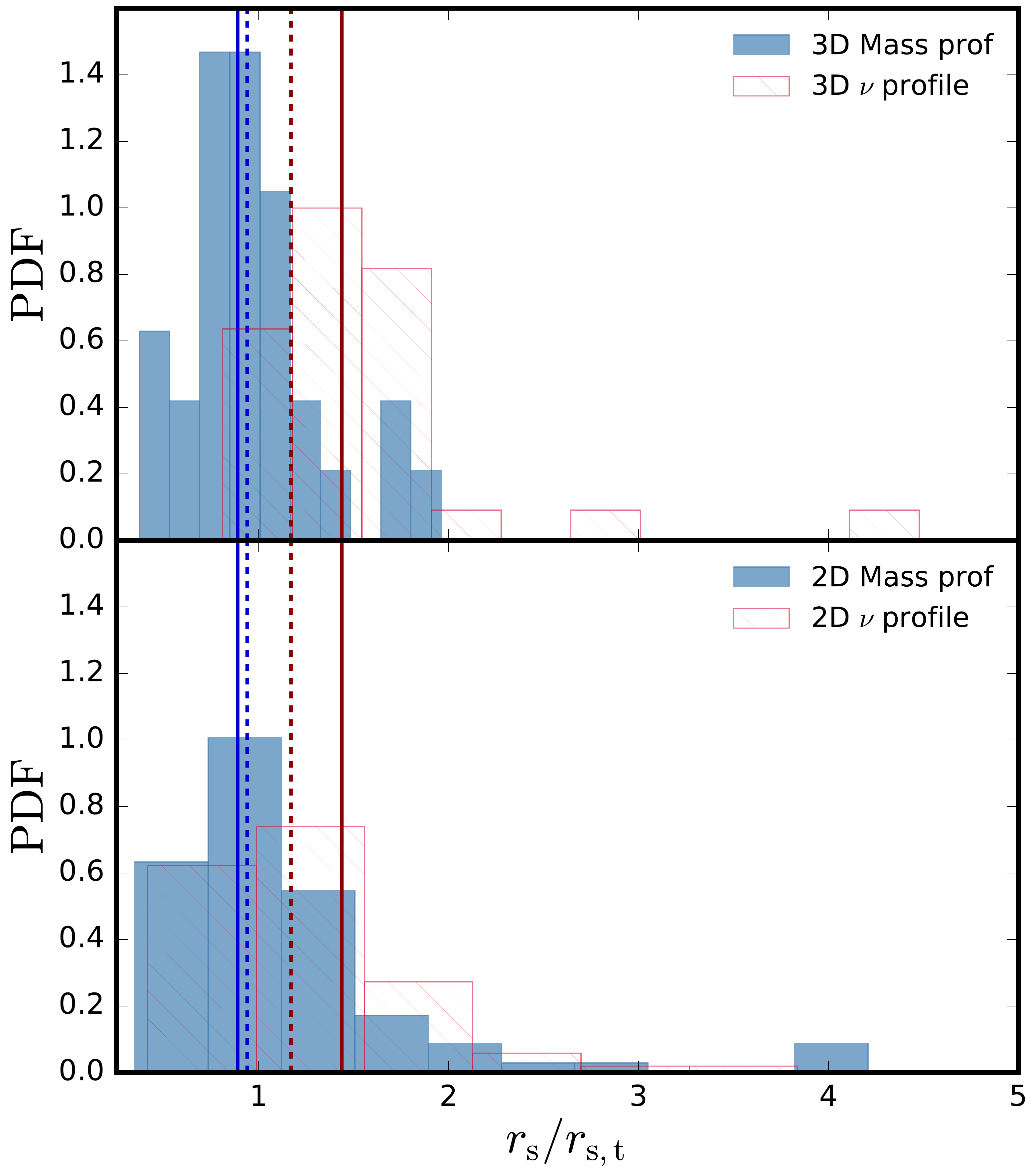}
		\caption{The ratio of the measured values of $r_\mathrm{s}$ for the galaxy number density and the recovered mass profile with respect to the $r_\mathrm{s}$ obtained from the true mass profile. The top and bottom panels show the $r_s$ values for the 3D and 2D profiles respectfully. The vertical bars shows the median ratio of each histogram with the corresponding colour, with the solid lines representing the 3D case and the dashed lines the 2D case. The histogram areas are normalised to one. We can see that while the scale radius for the overall mass profile is close to the true value for both the 3D and 2D cases there is a significant shift in the galaxy number density profile.}
		\label{fig:RsComparison}
	\end{figure}

	Here we show how the recovered values of $r_s$ and $p$ vary between the `3D' and `2D' cases. In Fig. \ref{fig:RsComparison} we show that the recovered values of $r_s$ for the number density profile differ significantly from the $r_s$ of both the true (i.e. $r_s$ from the particles) and recovered mass profiles. This implies that the galaxies are not fair tracers of the underlying density profile, and justifies the assumption that the two $r_s$ values should be fit independently.
	
\begin{figure}
		\centering
		\includegraphics[width=0.99\linewidth]{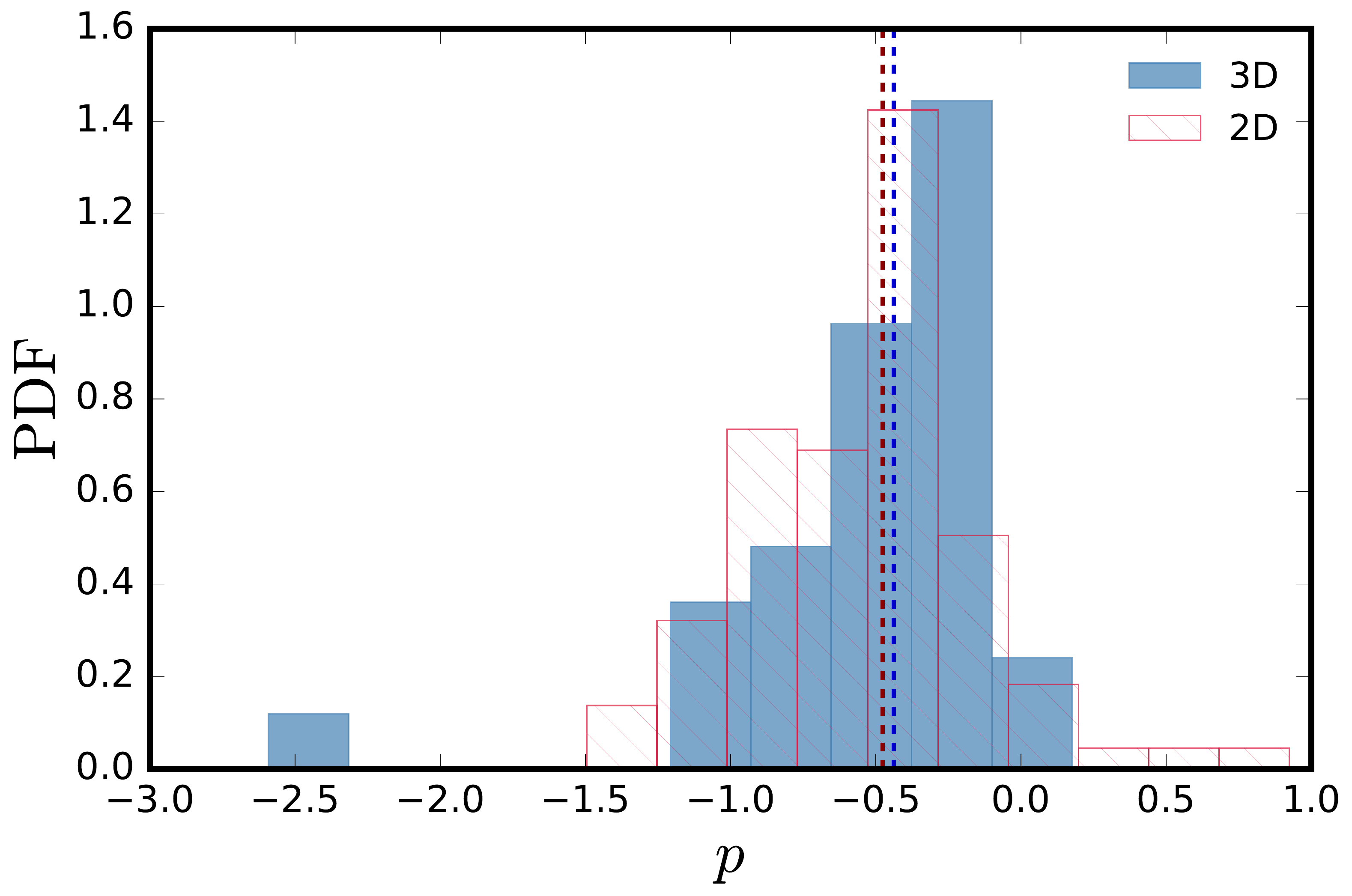}
		\caption{The distribution of $p$ in equation (\ref{sigma_plaw}) for the 3D (blue) and 2D (cyan) cases. The vertical bars show the median value of $p$ and the area under each histogram is normalised to one.}
		\label{fig:pwrHist}
	\end{figure}

	We show the distribution of the exponent $p$ from equation (\ref{sigma_plaw}) in Fig. \ref{fig:pwrHist}. The median values are similar for both the 3D and 2D cases at ${\sim} 0.5$. However, there are several instance where $p>0$, one of which is CE-05 when projected along the $z$ axis in Fig. \ref{fig:sigPWR26}. CE-06 is the cluster with $p=0.6 \pm 0.7$ in the 3D case. CE-06 is a highly disturbed cluster in the process of merging, containing two distinct cores.
	


	\bsp	
	\label{lastpage}
\end{document}